\documentclass[aps,prb,amsmath,twocolumn,amssymb,floatfixng,showpacs,
superscriptaddress,footinbib]{revtex4-1}
\pdfoutput=1
\usepackage[dvips]{graphics}
\usepackage{bm}
\usepackage{float}
\usepackage{epsfig}
\usepackage{enumerate}
\usepackage{color}
\usepackage{graphicx}
\usepackage[colorlinks=true,linktoc=page,linkcolor=magenta,citecolor=cyan]{hyperref}

\newcommand{\etal}{{\it et al.~}}
\newcommand{\ie}{{\it i.e., }}

\newcommand{\ua}{{\uparrow }}
\newcommand{\da}{{\downarrow }}
\newcommand{\eps}{{\epsilon}}
\newcommand{\la}{{\langle}}
\newcommand{\ra}{{\rangle}}
\newcommand{\suc}{{superconductor~}}

\begin{document} 
\title{Enhancement of crossed Andreev reflection in normal-superconductor-normal junction of thin topological insulator} 

\author{SK Firoz Islam}
\email{firoz@iopb.res.in}
\affiliation{Institute of Physics, Sachivalaya Marg, Bhubaneswar-751005, India}
\author{Paramita Dutta}
\email{paramitad@iopb.res.in}
\affiliation{Institute of Physics, Sachivalaya Marg, Bhubaneswar-751005, India}
\author{Arijit Saha}
\email{arijit@iopb.res.in}
\affiliation{Institute of Physics, Sachivalaya Marg, Bhubaneswar-751005, India}
\affiliation{Homi Bhabha National Institute, Training School Complex, Anushakti Nagar, Mumbai 400085, India}

\begin{abstract}
We theoretically investigate the subgapped transport phenomena through a normal-superconductor-normal (NSN) junction made up 
of ultra thin topological insulator with proximity induced superconductivity. The dimensional crossover from three dimensional 
($3$D) topological insulator (TI) to thin two-dimensional ($2$D) TI introduces a new degree of freedom, the so-called 
hybridization or coupling between the two surface states. We explore the role of hybridization in transport properties of 
the NSN junction, especially how it affects the crossed Andreev reflection (CAR). We observe that a rib-like pattern appears 
in CAR probability profile while examined as a function of angle of incidence and length of the superconductor. Depending 
on the incoming and reflection or transmission channel, CAR probability can be maneuvered to be higher than $97\%$ under
suitable coupling between the two TI surface states along with appropriate gate voltage and doping concentration in the normal region. 
Coupling between the two surfaces also induces an additional oscillation envelope in the behavior of the angle averaged conductance, with the 
variation of the length of the superconductor. The behavior of co-tunneling (CT) probability is also very sensitive to the coupling and other parameters. 
Finally, we also explore the shot noise cross correlation and show that the behavior of the same can be monotonic or non-monotonic depending on the 
doping concentration in the normal region. Under suitable circumstances, shot noise cross correlation can change sign from positive to negative or 
vice versa depending on the relative strength of CT and CAR.
\end{abstract}

\maketitle
\section{Introduction}\label{sec1}
The phenomenon of electron-hole conversion across the interface of a normal metal and superconductor, known as Andreev 
reflection~\cite{Andreev1}(AR), has been paid much attention in last decade especially after the pioneering work by C. 
W. Beenakker~\cite{beenakker2006specular} revealing the unusual specular AR in an undoped graphene~\cite{neto2009electronic}. 
The origin behind this intriguing specular AR in the graphene lies in the low energy gapless linear band dispersion which stems from 
its hexagonal lattice geometry. In AR process, an incident electron from the normal metal with energy less than the superconducting gap forms 
a Cooper pair of charge $2e$, $e$ being the electronic charge, inside the \suc leaving behind a hole with opposite spin in the 
normal metal region~\cite{PhysRev.106.162}.

On the other hand, crossed Andreev reflection (CAR)~\cite{falci2001correlated,bignon2004current,recher2001andreev,PhysRevB.74.214510,pauldutta} 
is another intriguing phenomenon appearing in a normal-superconductor-normal (NSN) hybrid junction where the superconducting
length ($L$) is comparable to the coherence length ($\xi$) of the Cooper pair. In CAR process, an incident electron from one of the 
normal metal regions together with another electron of opposite spin forms a Cooper pair leaving a hole into the other 
normal side. A series of experiments~\cite{russo2005experimental,chandrasekhar,WeiChandrasekhar,AndyDas,LHofstetter,
LGHerrmann,ZimanskyChandrasekhar,JBrauer} have been reported realizing CAR phenomenon. One of the major applications 
of CAR process is to generate entangled electron pairs by breaking the Cooper pair through two spatially separated metallic leads 
attached to a superconductor known as beam splitter~\cite{recher,WeiChandrasekhar,veldhorst2010nonlocal,samuelsonbuttiker,LesovikMatin,yeyati,PhysRevB.78.235403}.
Possibility of applications of CAR process has invoked researchers to propose several ways to enhance CAR in various 
materials~\cite{beckmann2004evidence,yamashita2003crossed,weichen1,russo2005experimental,soori2016enhancement}. 

Note that, the CAR is associated with another competitive quantum mechanical scattering process known as elastic co-tunneling (CT) of electron. In recent 
past, J. Cayssol~\cite{cayssol2008crossed} has shown that the AR and the CT can be completely suppressed by suitably choosing the doping level in the 
normal region of a graphene ($n$-type)-superconductor-graphene ($p$-type) heterostructure, leading towards the first step of possible realization 
of entanglement in Dirac material. In recent times, several theoretical works of spin selective CAR phenomena have been carried out in 
silicene~\cite{paulquantum} and transition-metal dichalkogenides material-MoS$_2$~\cite{majidi2014valley}.


On the other hand, very strong spin-orbit interaction may lead to conducting surface states associated with insulating 
bulk in some materials known as topological insulator (TI)~\cite{PhysRevLett.95.146802,bernevig2006quantum,konig2007quantum,
PhysRevB.75.121306,PhysRevLett.98.106803,zhang2009topological}. It is the time-reversal symmetry, inherited by materials 
like Bi$_{2}$Se$_{3}$, Sb$_2$Te$_3$ and Bi$_2$Te$_3$~\cite{zhang2009topological} etc. protecting this unique feature. 
Though experimental realization of conducting surface states in 3D topological insulators (TIs) has been reported by several 
groups~\cite{hsieh2008topological,Hsieh919,xia2009observation,roushan2009topological}, one of the major obstacles is to 
isolate the transport properties of the surface states from the unavoidable bulk contribution. 
This problem has been resolved by growing the TI sample in the form of ultra-thin film~\cite{zhang2009quintuple,peng2010aharonov,zhang2010crossover}, 
in which bulk contribution becomes vanishingly small. The small thickness in thin TI favors the overlapping between the top and bottom 
surface states introducing a new degree of freedom which is coupling or hybridization between the two surface states. However, it 
is limited to a certain thickness of five to ten quintuple layers which is of the order of $10$ nm~\cite{zhang2010crossover, PhysRevLett.108.216803}. 

Recently, several experimental realizations of proximity induced superconductivity in TI~\cite{PhysRevB.84.165120,sacepe2011gate,
veldhorst2012josephson,PhysRevX.4.041022,PhysRevLett.117.147001,wang2012coexistence}, as well as theoretical 
investigations of AR phenomena in TI have been carried out~\cite{PhysRevB.87.245435,niu2010crossed,PhysRevB.93.195404, 
PhysRevB.82.115312,weichen2}. However, Majidi \etal~\cite{PhysRevB.93.195404}, have shown that the coupling between the top and bottom 
surface of an ultra-thin TI can lead to the intra-band specular AR which is in complete contrast to graphene in which 
specular AR is the inter-band type~\cite{beenakker2006specular}. In addition to this, it has been pointed out that AR with $100\%$ probability 
can be achieved for a wide range of angle of incidence under suitable circumstances. On the contrary, in graphene it happens only for normal angle of 
incidence~\cite{beenakker2006specular}. The concept of intra-band specular Andreev reflection was first put forwarded by Bo Lv \etal~\cite{PhysRevLett.108.077002} 
in usual $2$D electron gas with strong Rashba spin-orbit interaction, which has been exploited in thin TI~\cite{PhysRevB.93.195404}. 
Note that, in $2$D quantum spin Hall systems, CAR is found to be completely suppressed~\cite{PhysRevB.82.081303,weichen2}, if there is no
coupling between the two edges.

The interesting signatures of the coupling in AR phenomenon~\cite{PhysRevB.93.195404} have motivated us to carry out a meticulous study of the 
CAR in a NSN hybrid junction, especially to reveal the role of coupling in  the scattering processes and conductance, shot noise therein.
We consider a thin TI ($n$-type)-superconductor-thin TI ($p$-type) heterostructure and investigate the CAR, CT and conductance by using
the extended Blonder-Tinkham-Klapwijk (BTK) formalism~\cite{BTK}. We observe that hybridization between the two surface states can 
enhance CAR up to $97\%$ even when the normal regions are sufficiently doped. The remaining $3\%$ is the normal reflection probability.
This results in enhancement of conductance  due to the CAR process too. Additionally, coupling induces a weak oscillation in the CAR 
conductance with the length of the superconductor. Note that, in our proposed model, we consider the NSN hybrid junction where superconducting
correlation is induced in thin TI by placing it in close proximity to a bulk superconductor~\cite{PhysRevLett.117.147001,wang2012coexistence}. 
We also investigate the shot noise cross-correlation for the transport phenomena in our model hybrid junction. In normal metal-superconductor 
hybrid junction, shot noise has some important diagnostic features like detecting open transmission channel~\cite{beenakker2006quantum} 
and entanglement~\cite{samuelsonbuttiker,LesovikMatin,yeyati,PhysRevB.78.235403,PhysRevB.61.R16303,PhysRevB.61.R16303,WeiChandrasekhar,AndyDas} etc. 
Recently, shot noise measurement has been successfully carried out in Dirac material
like graphene~\cite{PhysRevLett.100.196802}. We show that within suitable parameter regime, shot noise cross-correlation exhibits positive sign indicating the existence of 
possible entangled states~\cite{samuelsonbuttiker,LesovikMatin,yeyati}.

The rest of the paper is structured as follows. The model Hamiltonian and energy dispersion of each region have been discussed 
in Sec.~\ref{sec2}. In Sec.~\ref{sec3}, we present our numerical results for scattering amplitudes, conductance and shot noise 
cross correlation for the NSN hybrid structure of thin TI. Finally, we summarize and conclude in Sec.~\ref{sec4}.

\section{Model Hamiltonian and energy dispersion}\label{sec2}
In this section, we discuss the model Hamiltonian and corresponding energy dispersion of different regions of our NSN hybrid 
\begin{figure}[!thpb]
\centering
\includegraphics[height=4cm,width=0.90 \linewidth]{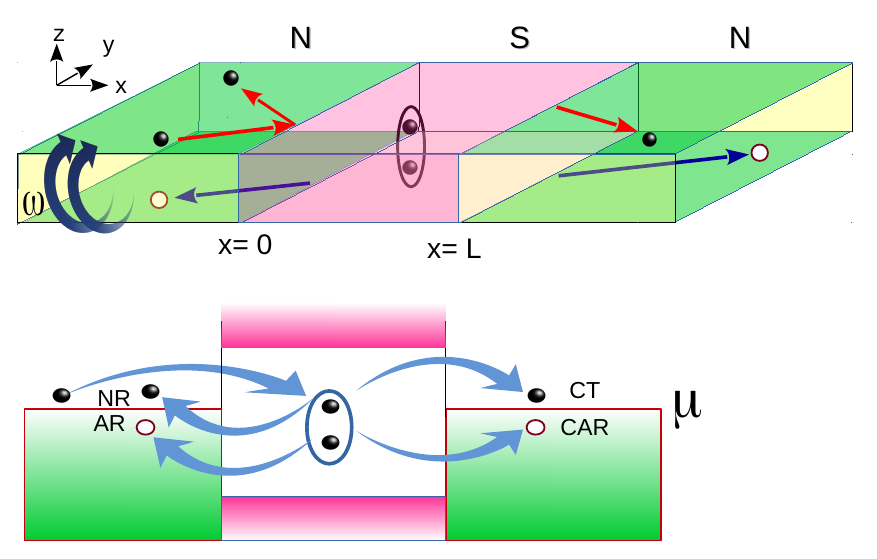}
\caption{(Color online) Schematic diagram of our geometry and various scattering phenomena, occurring at the thin-topological insulator 
NSN hybrid structure, is displayed. Light green (light grey) and pink (grey) shadowed regions correspond to the normal (N) and superconducting (S) regions, respectively. 
Upper panel: Red (black) and blue (black) arrows indicate the direction of electron (black solid bullet) and hole (white hollow bullet), respectively along with the Cooper pair 
inside the S-region of length $L$. The coupling between the top and bottom surfaces is denoted by $\omega$. Lower panel: Different scattering phenomena in the context 
of energy band is illustrated considering proximity induced effective pairing gap in S-region. $\mu$ denotes the chemical potential.}
\label{Fig1}
\end{figure}
junction of thin TI, following Ref.~[\onlinecite{PhysRevB.93.195404}], for normal-superconductor (NS) structure. We consider that the 
entire system lies in the $x$-$y$ plane and a perpendicular electric field is applied along the $z$-direction between the two surfaces via 
gate electrodes ($U_{top}=U$ and $U_{bottom}=-U$). The middle region of the thin film, as shown by pink shadow in 
Fig.~\ref{Fig1}, is the proximity induced superconducting region. The pairing between electron and hole via superconductor 
can be expressed by the Dirac-Bogoliubov-de Gennes (DBdG) equation as,
\begin{equation}
\left[\begin{array}[c]{c c}
            \hat{H}(p)-\mu & \hat{\Delta}_S\\
            -\hat{\Delta}^{\ast}_{S} &\mu-\hat{H}^{\ast}(-p) \\
            \end{array}\right]\left[\begin{array}[c]{c}u\\v\end{array}\right]
            =\epsilon\left[\begin{array}[c]{c}u\\v\end{array}\right].
\end{equation}
Here, $\mu$ is the chemical potential and $\Delta_S$ is the proximity induced superconducting pair potential. The effective single particle 
Hamiltonian of the thin topological insulator ~\cite{PhysRevB.83.245428,PhysRevB.92.045429,PhysRevB.86.165404} can be written as
(see Appendix \ref{App} for the derivation)
\begin{equation}
 \hat{H}(p)=\hat{\tau}_{z}\otimes\hat{h}(p)+\hat{\tau}_x\otimes\omega\hat{\sigma}_0+U\hat{\tau}_z\otimes\hat{\sigma}_0 \ ,
 \end{equation}
which acts on four component eigen states $\Psi=[\psi_{\it t}^{\ua}, \psi_{\it t}^{\da}, \psi_{\it b}^{\ua}, \psi_{\it b}^{\da}]^{T}$.
Here $\omega$ is the coupling between the top and bottom surface states. The energy of the incident electron is denoted by $\eps$. The pairing 
symmetry inside the superconducting region is considered to be inter-surface $s$-wave as used in Ref.~[\onlinecite{PhysRevB.93.195404}]. It is given by 
$\hat{\Delta}_S=i\Delta_S\hat{\sigma}_y[\Theta(x-L)-\Theta(x)]$ where $\Theta$ is the Heaviside step function. 
In the low energy regime, we can write
\begin{equation}
\hat{h}(p)=v_{F}(\hat{\sigma}\times\hat{p})_{z} 
\end{equation}
with ${\bf p}=\{p_x,p_y\}$ being the $2$D momentum operator and $v_{F}$ denoting the Fermi velocity. We consider here
the low energy approximation of $\hat{h}(p)$. With the reduction of thickness of TI, the strength of the coupling between 
top and bottom surface states ($\omega$) is enhanced. The two sets of Pauli matrices \ie $\hat{\sigma}$ and $\hat{\tau}$
act on real spin and surface pseudo spin degree of freedom. The energy spectrum in the normal region is given by
\begin{equation}\label{dis}
 \eps_{\lambda}^{\pm}(k)=\lambda\sqrt{(\hbar v_F |k| \pm U)^2+\omega^2}-\mu_{N} \ .
\end{equation}
where $\lambda=\pm$ denotes band index and $\mu_N$ is the chemical potential in the normal region. The coupling parameter 
$\omega$ controls the band gap between the two surface bands. On the other hand, it is the gate voltage ($U$) which causes 
energy splitting in the same band. 

Similarly inside the superconducting region, characterized by chemical potential $\mu_S$, energy dispersion is given by
\begin{equation}\label{band_S}
 \eps^{\pm}(k_S)=\sqrt{\mu_S-\sqrt{(\hbar v_F|k_S|\mp U)^2+\omega^2}+\Delta_S^2} \ .
\end{equation}
The incident electron can undergo four possible scattering events. It can either be normally reflected 
as an electron via normal reflection (NR), Andreev reflected as a hole with opposite spin, or be transmitted as an electron via CT 
and as a hole with opposite spin via CAR. These four scattering processes are schematically displayed in the lower panel of Fig.~\ref{Fig1}. 
Note that, unlike graphene where Andreev phenomenon occurs between two valleys, here each conducting surface contains single Dirac cone.
Hence Andreev phenomenon occurs between the two surfaces (top and bottom) due to the inter-surface pairing symmetry as assumed in our analysis.
If the incident electron belongs to the top surface, then the reflected or transmitted hole via the AR and CAR respectively take place 
in the bottom surface. On the other hand, NR and CT correspond to the top surface as shown in the upper panel of Fig.~\ref{Fig1}.

\begin{figure}[!thpb]
\centering
\includegraphics[height=4.0cm,width=0.80 \linewidth]{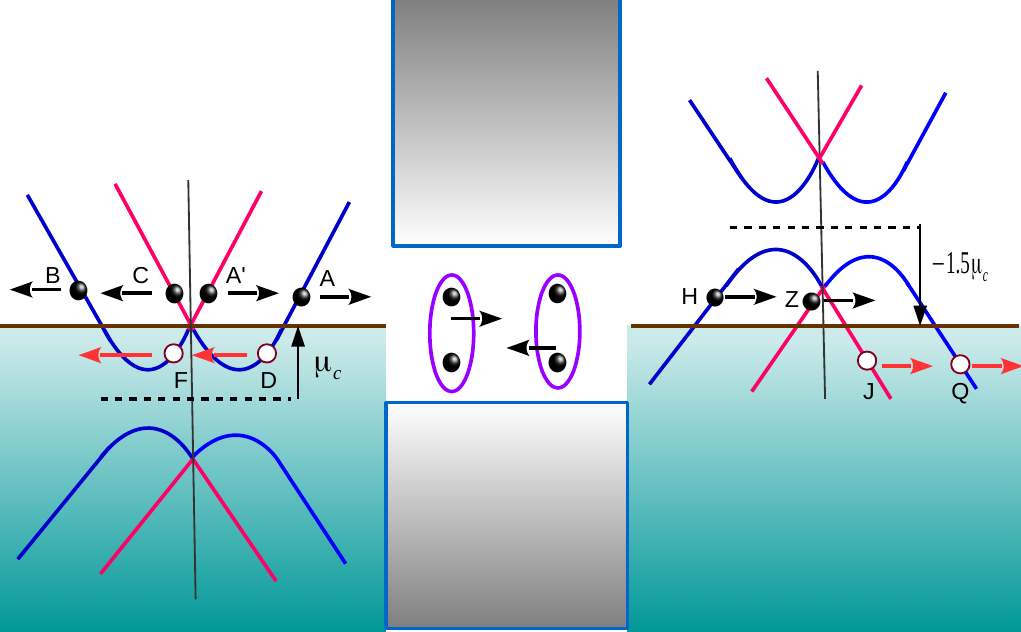}
\caption{(Color online) Schematic diagram of different scattering processes for the chemical potential $\mu_L=\mu_c$ and 
$\mu_R=-1.5 \mu_L$ is depicted. The left and right regions are the normal thin TI while the middle one is the proximity induced 
superconducting region following the energy dispersion as given in Eqs.(\ref{dis})-(\ref{band_S}). Electrons and holes are denoted by 
black bullets and empty white bubbles, respectively. The horizontal black arrows indicate the direction of propagation 
of electrons while, the red (light grey) ones are used for the holes, respectively. Two energy branches $\eps_{\lambda}^{+}$ and 
$\eps_{\lambda}^{-}$ are indicated by red (grey) and blue (grey) curves, respectively.}
\label{Fig2}
\end{figure}

\section{Numerical results}\label{sec3}
In this section we present our numerical results for the scattering amplitudes, conductance and shot-noise in three different sub-sections. We discuss our results in terms of the scattering processes occurring at the interface of the hybrid structure and various parameters of the system.
\subsection{Scattering amplitudes} 
In order to discuss the results for the scattering amplitudes, we consider two different situations by setting the chemical potential in the right normal thin TI region at two different doping levels ($p$-type). 
They are at $\mu_R=-1.5 \mu_L$ and 
$\mu_R=-\mu_L$ respectively while the chemical potential $\mu_L$ at the left normal thin TI region 
is fixed to $\mu_c$ ($n$-type). Here, $\mu_c=\sqrt{U^2+\omega^2}$ 
is the critical chemical potential at which energy branches cross each other at zero momentum. 
The \suc is doped at $\mu_{S}=1$ eV. We choose $\mu_{S}\gg\Delta_{S}$ for the requirement of the mean-field treatment of superconductivity~\cite{beenakker2006specular,beenakkerreview}.
\subsubsection{Case I : $\mu_{_R}=-1.5\mu_{_L}$}

In Fig.~\ref{Fig2} we show a schematic diagram for the different scattering channels. The gate voltage induced energy splitting in the same band opens up two different incident channels ($A$ and $A'$) 
for  an electron with the same energy but with different momentum. An electron with incoming energy $\epsilon>0$,  incident from $A$ or $A'$, can either be reflected or transmitted as an electron 
or hole through a pair of reflection and transmission channels. The normal reflection as electron can happen through $ B$ and $C$ whereas the possible Andreev 
reflection channels are $D$ and $F$ respectively. On the other hand, transmission either as an electron via CT or as a hole via CAR can take place 
via $ Z$ and $ H$ or $ J$ and $ Q$, respectively, as demonstrated in Fig.~\ref{Fig2}. The Andreev reflection 
corresponding to $D$ channel is retro type while it is specular for the channel $F$~\cite{PhysRevB.93.195404}.
Most remarkably, this specular AR is intra-band which is in complete contrast to graphene where inter-band specular AR
was predicted by Beenakker~\cite{beenakker2006specular}. However, in thin TI, CT and CAR can be either intra-branch or 
inter-branch type originating from the same band. 

To obtain the different scattering amplitudes, we match the wave functions across the boundary at $x=0$ and $x=L$ (see Fig.~\ref{Fig1}), 
\ie $\Psi_{L}|_{x=0}=\Psi_{S}|_{x=0}$ and $\Psi_{S}|_{x=L}=\Psi_R|_{x=L}$ where $\Psi_{L(R)}$ and $\Psi_{S}$ are the wave 
functions corresponding to the left (right) normal region and superconducting region, respectively (see 
Appendix~\ref{App1} for the explicit form of the wave functions). For an incoming electron from $j=\{A,A'\}$, we denote 
the CT and CAR amplitudes by $t_Z^{e}$, $t_H^{e}$ and $t_Q^{h}$, $t_J^{h}$ corresponding to the channels $Z$, $H$, $Q$ 
and $J$, respectively. On the other hand, NR and AR amplitudes are denoted by $r_B^{e}$, $r_C^{e}$ and $r_D^{h}$, $r_F^{h}$ for the 
channels $B$, $C$, $D$, and $F$ respectively. The superscript `$e$' and `$h$' denote electron and hole, respectively. 
Note that, for both the incoming channels ($A$ and $A^{\prime}$) scattering probabilities satisfy unitarity condition. It can be expressed as
\begin{equation} 
\sum_{\iota} R_{\iota}^{e}+\sum_{\eta} R_{\eta}^{h}+\sum_{\bar{\iota}} T_{\bar{\iota}}^{e}+\sum_{\bar{\eta}} 
T_{\bar{\eta}}^{h}=1\ .
\end{equation}
where $R_{\iota}^{e}=|r_{\iota}^{e}|^2$ with $\iota=\{B,C\}$, $R_{\eta}^{h}=|r_{\eta}^{h}|^2$ with $\eta=\{F,D\}$, 
$T_{\bar{\iota}}^{e}=|t_{\bar{\iota}}^{e}|^2$ with ${\bar{\iota}}=\{Z,H\}$ and 
$T_{\bar{\eta}}^{h}=|t_{\bar{\eta}}^{h}|^2$ with ${\bar{\eta}}=\{J,Q\}$.  The wave vectors for different
transmission channels are given by $|k_{Z(H)}|=[\sqrt{(\eps+\mu_R)^2-\omega^2}\mp U]/(\hbar v_{F})$ and 
$|k_{J(Q)}|=[\sqrt{(\eps-\mu_R)^2-\omega^2}\mp U]/(\hbar v_{F})$.
 \begin{figure}[htb]
\begin{minipage}[t]{0.5\textwidth}
  \hspace{-.2cm}{ \includegraphics[width=.5\textwidth,height=4cm]{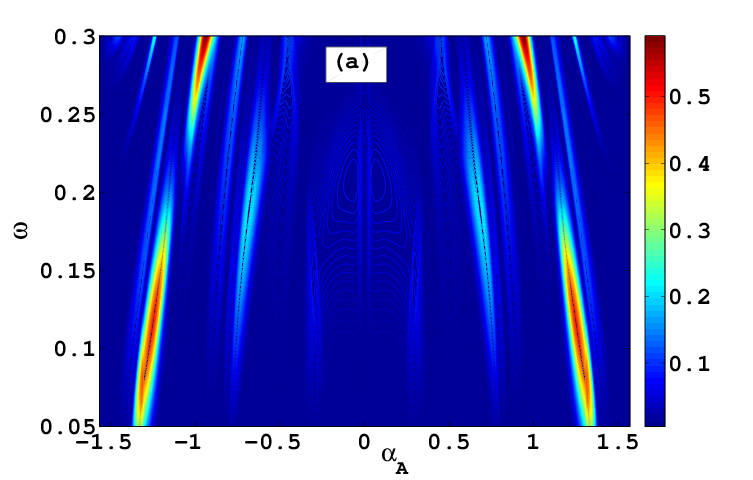}}
  \hspace{-.14cm}{ \includegraphics[width=.5\textwidth,height=4cm]{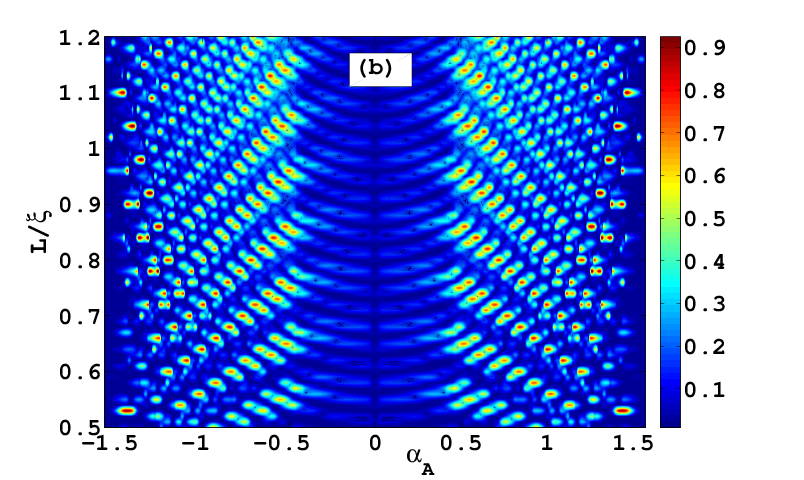}}
\end{minipage}
\begin{minipage}[t]{0.5\textwidth}
  \hspace{-.2cm}{ \includegraphics[width=.5\textwidth,height=4cm]{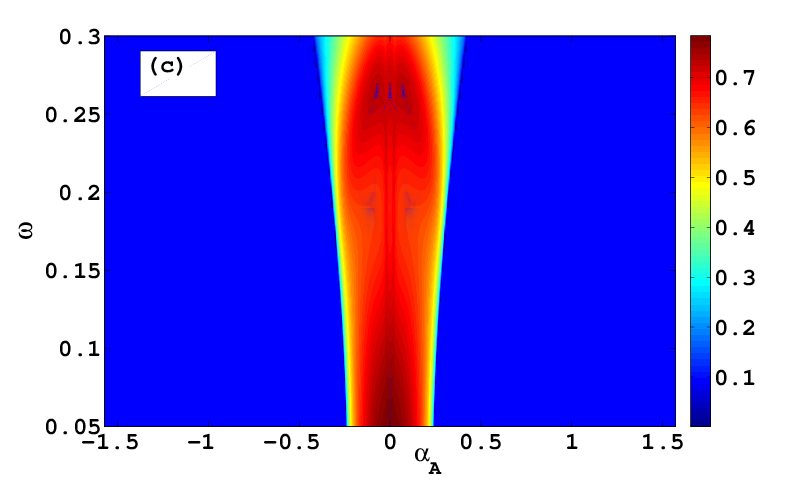}}
  \hspace{-.3cm}{ \includegraphics[width=.5\textwidth,height=4cm]{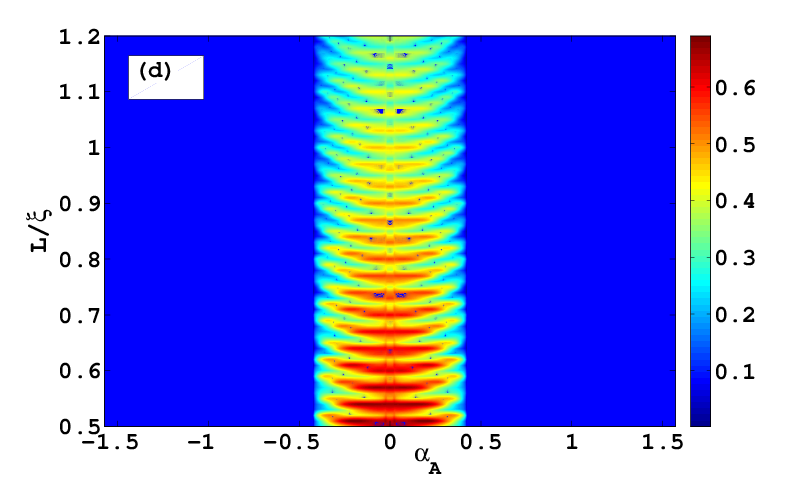}}
\end{minipage}
\caption{(Color online) The behavior of CAR probability at $Q$ \ie $T_{Q}^{h}$ is illustrated in panel (a) $\omega$$-$$\alpha_A$ plane
and panel (b) $L/\xi$ $-$ $\alpha_A$ plane for an incident electron at $A$. Similarly, CT probability at $Z$ ($T_Z^{e}$) is shown
in panel (c) and (d) in the same parameter space. The value of the other parameters are chosen to be gate potential $U=0.3$ eV, excitation energy $\eps/\Delta_S=1$. 
We choose $L/\xi=0.5$ for the left column and $\omega=0.3$ eV for the right column.}
 \label{Fig3}
 \end{figure}

For the above-mentioned scenario, we discuss the features of CT and CAR probabilities as a function of the coupling strength $\omega$, 
angle of incidence $\alpha_A$ and the length $L$ of the superconducting region. We set the energy of the incident electron as $\eps/\Delta_S=1$. 
In Ref.~[\onlinecite{PhysRevB.93.195404}] it is already been explored that specular AR can occur with $100\%$ efficiency for a 
wide range of angle of incidence at this energy. Hence, we also choose this particular energy value for our investigation of CAR. 
In the left column of Fig.~\ref{Fig3}, we show the behavior of CAR and CT probability in the $\omega$$-$$\alpha_A$ plane, while in the right column 
the same has been depicted in the $L/\xi$$-$$\alpha_A$ plane, for an incoming electron from channel $A$. From Fig.~\ref{Fig3}(a), 
we observe that CAR probability at $Q$ ($T_{Q}^{h}$) exhibits a maxima for a particular angle of incidence $\alpha_A$ and coupling constant $\omega$. 
 $T_{Q}^{h}$ can reach around $50\%$ probability for the set of parameter values like $\alpha_A\simeq\pm0.95$ and $\omega=0.3$ eV
as well as $\alpha_A=\pm1.3$ and $\omega \sim 0.1$ eV. Note that, this CAR process is inter-band but intra-branch type with 
the incident electron and transmitted hole energy as $\eps^{-}_{+}$ and $\eps^{-}_{-}$, respectively. This feature is
shown for $L/\xi=0.5$. In Fig.~\ref{Fig3}(b), we investigate the behavior of CAR as a function of the length of the superconducting region $L/\xi$ and
find that it exhibits a rib-like pattern characterized by several resonances with the variation of both the length of 
the superconducting region and angle of incidence $\alpha_A$. Note that, in the $L/\xi$$-$$\alpha_A$ plane, CAR can be achieved even with $90\%$
probability under suitable circumstances. However, it is found to be absent for normal incidence ($\alpha_A=0$) as in this case all the electrons 
are locally reflected as holes due to AR. 

On the other hand, CT manifests a completely different behavior with the variation of the length of S-region as well as the angle of incidence.
 \begin{figure}[htb]
 \begin{minipage}[t]{0.5\textwidth}
  \hspace{-.1cm}{ \includegraphics[width=.5\textwidth,height=4cm]{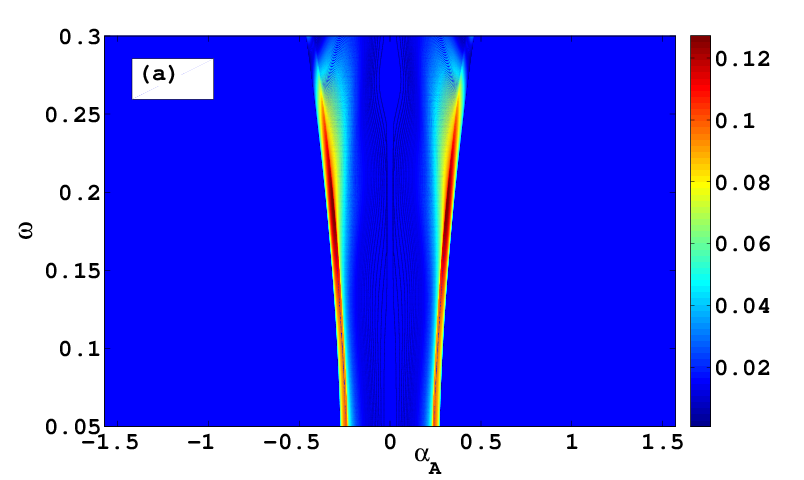}}
  \hspace{-.2cm}{ \includegraphics[width=.5\textwidth,height=4cm]{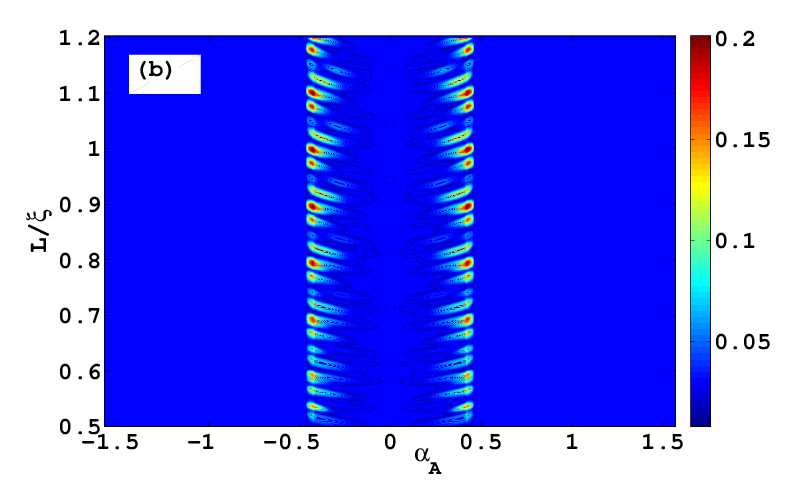}}
\end{minipage}
\begin{minipage}[t]{0.5\textwidth}
  \hspace{-.1cm}{ \includegraphics[width=.5\textwidth,height=4cm]{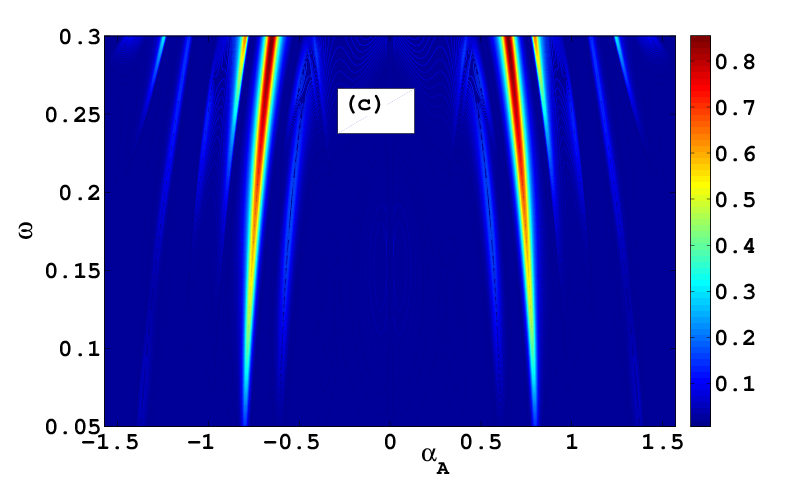}}
  \hspace{-.2cm}{ \includegraphics[width=.5\textwidth,height=4cm]{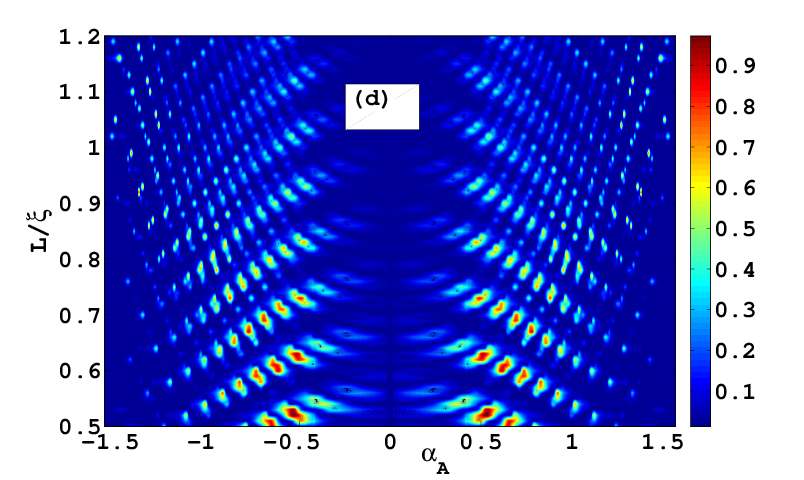}}
\end{minipage}
\caption{(Color online) The behavior of CAR probability at $J$ \ie $T_{J}^{h}$ is shown in panel (a) and (b) in the 
$\omega$ $-$ $\alpha_A$ and $L/\xi$ $-$ $\alpha_A$ plane, respectively. Similarly, the behavior of CT probability at $H$ ($T_H^{e}$)
in the plane of $(\omega,\alpha_A)$ and ($L/\xi,\alpha_A$) are portrayed in panel (c) and (d), respectively. The value of the other parameters
are chosen to be the same as mentioned in Fig.~\ref{Fig3}.}
\label{Fig4}
\end{figure}
In Fig.~\ref{Fig3}(c), we observe that the probability of CT at $Z$ \ie $T_{Z}^{e}$ exhibits a continuous band-like profile 
around an angular region, confined by the critical angle $\alpha_A^{c}$, as one increases the coupling strength $\omega$
between the two surfaces. Most interestingly, maxima of $T_Z^{e}$ appears even for relatively weak coupling strength in contrast to CAR at $Q$
as it's an intra-band process. Although $T_Z^{e}$ is inter-branch type as incident and transmitted electrons are from $\eps_{+}^{-}$ and $\eps_{-}^{+}$,
respectively. It can be explained as Klein tunneling phenomena as $T_Z^{e}=1$ at $\alpha_{A}=0$, being almost independent of the coupling strength.
When we change the length of the superconducting region, the behavior of CT manifests an oscillatory pattern as displayed in Fig.~\ref{Fig3}(d). 
It mimicks a spinal-chord like pattern with a linearly decaying amplitude with the enhancement of the superconducting length within the 
range confined by $-\alpha_A^c$ to $\alpha_A^c$. It is apparent from Fig.~\ref{Fig3}(c)-(d) that  the CT phenomena at $Z$ is limited by the critical angle 
in contrast to the CAR at $Q$ channel.

In Fig.~\ref{Fig4}, we discuss the behavior of CAR and CT probabilities for other transmission channels, \ie CAR at $J$ and CT at $H$ 
(see Fig.~\ref{Fig2}). In Fig.~\ref{Fig4}(a) we show the features of $T^h_J$ in the $\omega$$-$$\alpha_A$ plane for a fixed value 
of $L$ ($=0.5 \xi$). Throughout the contour CAR probability is vanishingly small except for a very narrow region on both sides of 
  \begin{figure}[htb]
   \begin{minipage}[t]{0.5\textwidth}
  \hspace{-.1cm}{ \includegraphics[width=.5\textwidth,height=4cm]{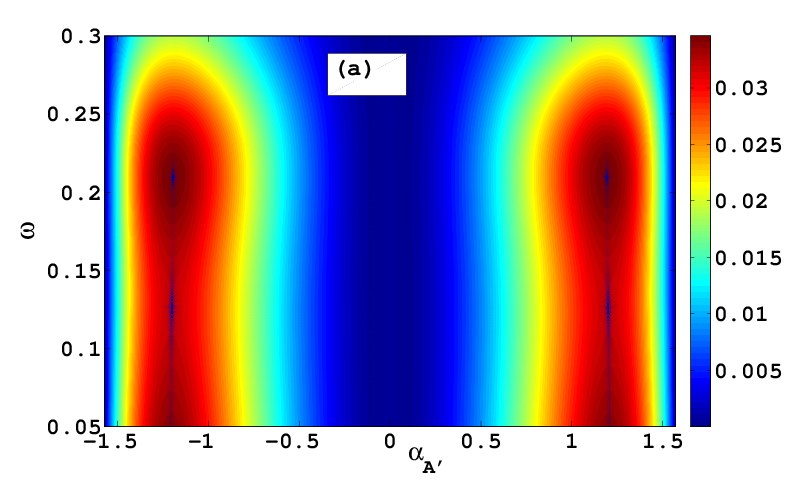}}
  \hspace{-.2cm}{ \includegraphics[width=.5\textwidth,height=4cm]{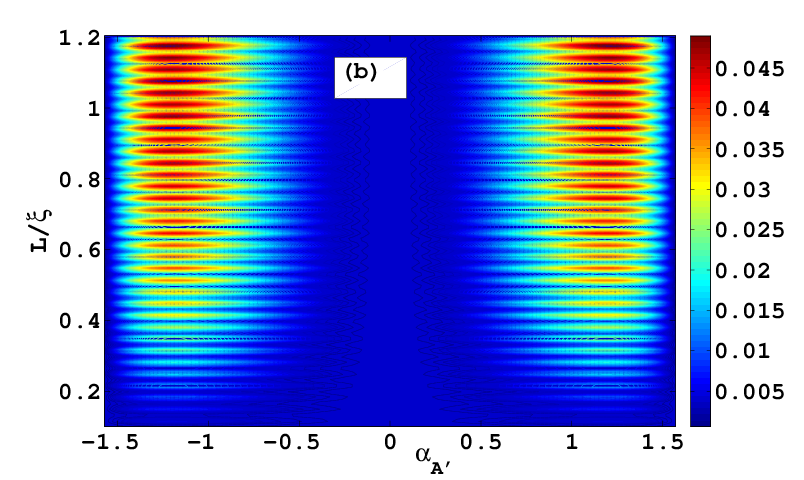}}
\end{minipage}
\begin{minipage}[t]{0.5\textwidth}
  \hspace{-.1cm}{ \includegraphics[width=.5\textwidth,height=4cm]{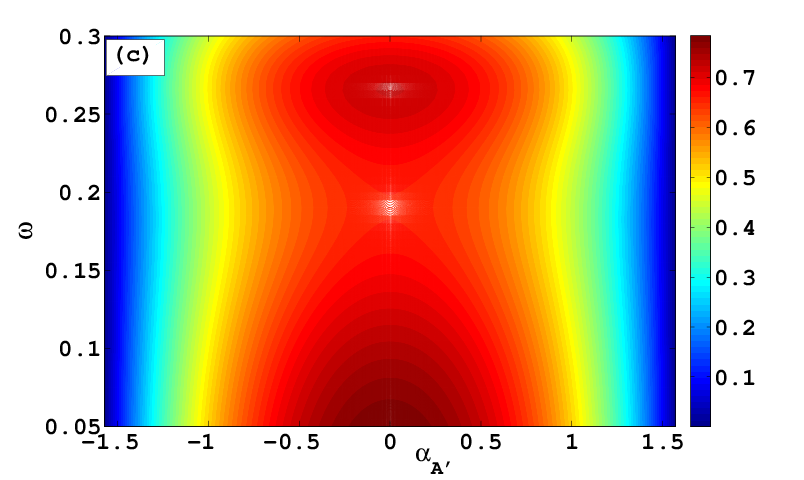}}
  \hspace{-.2cm}{ \includegraphics[width=.5\textwidth,height=4cm]{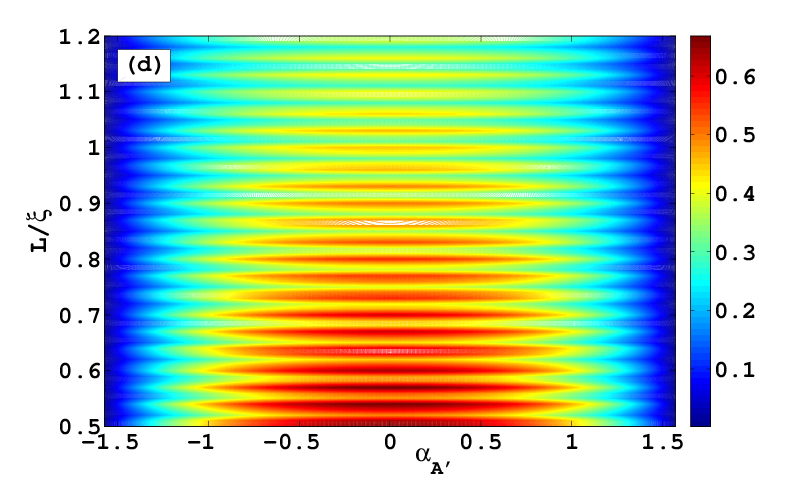}}
\end{minipage}
\caption{(Color online) The variation of CAR probability at $Q$, $T_{Q}^{h}$, is demonstrated in panel (a) and (b) in the 
plane of $(\omega,\alpha_{A'})$ and $(L/\xi,\alpha_{A'})$, respectively. Whereas, the behavior of CT probability at $Z$,
$T_Z^{e}$ is shown in the plane of $(\omega,\alpha_{A'})$ and ($L/\xi,\alpha_{A'}$) in panel (c) and (d),  respectively. 
Here the incident electron is considered via $A^{\prime}$ channel (see Fig.~\ref{Fig2}). We choose the same values of the other parameters 
as mentioned in Fig.~\ref{Fig3}.}
 \label{Fig5}
 \end{figure}
$\alpha_A=0$. The maximum probability of CAR via channel $J$, which arises for a regime of coupling $\omega \sim 0.15-0.25$, is of the order of $12\%$.
This is smaller in magnitude compared to the CAR obtained via channel $Q$. Such difference between the probabilities of CAR via the two 
channels can be explained as follows. For the electron incident from the channel $A$, change of energy branch associated 
with a momentum transfer, is required in order to obtain CAR at $J$. Whereas it is intra-branch process for $Q$. However, as one
increases the length of the  superconducting region, reduction or enhancement in CAR takes place at $J$ depending on the value of 
$\omega$ and $\alpha_A$. By varying the length of the superconducting region, we can achieve CAR with maximum $20\%$ probability being confined 
by the same angular space defined by the critical angle $\alpha_A^{c}$, as shown in Fig.~\ref{Fig4}(b). The latter manifests a
spinal chord-like pattern, in the behavior of CAR probability, with the variation of $L/\xi$ and $\alpha_A$.

On the other hand, the CT at $H$ ($T^{e}_{H}$) exhibits intriguing features as depicted in Fig.~\ref{Fig4}(c)-(d).
The behavior of CT is not limited by the critical angle as before. In Fig.~\ref{Fig4}(c) we show the behavior of $T^e_H$ 
in the $\omega$ $-$ $\alpha_A$ plane for $L/\xi=0.5$. Here, $T^e_H$ with maximum probability around $85\%$ can be obtained
within a very narrow region at a particular angle of incidence $\alpha_A$. Moreover, it increases with the strength 
of the coupling between the two surfaces of TI. However, the pattern changes to rib-like while we  investigate CT at $H$, 
in the plane spanned by $L/\xi$ and $\alpha_A$ as shown in Fig.~\ref{Fig4}(d). 
Also, it becomes oscillatory with decaying magnitude with the enhancement of the length of the superconductor for a fixed 
value of $\omega$ and $\alpha_A$. This feature is evident from Fig.~\ref{Fig4}(d). Similar behavior is obtained for some other values of $\alpha_A$ also. 
The decaying nature of CT is also obtained when we vary $\alpha_A$ for a particular value of $L/\xi$. Here, CT at normal incidence is found to 
be absent. Interestingly, more than $80\%$ inter-branch CT at $Z$ can be obtained for a wide range of angle of incidence while 
intra-branch CT at $H$ appears at a particular angle of incidence $\alpha_{A}$.

As mentioned earlier, for a particular energy there are two channels corresponding to two different momenta available 
for the electron to be incident on the NS interface. Here, we present our discussion of CAR and CT probabilities for an incoming electron 
incident from $A'$. In Fig.~\ref{Fig5}(a)-(b), we demonstrate the behavior of $T^h_Q$  in the plane of ($\omega$, $\alpha_{A^{\prime}}$)
and ($L/\xi$, $\alpha_{A^{\prime}}$) respectively. Similarly, Fig.~\ref{Fig5}(c)-(d) illustrate the variation of $T^e_Z$ in 
$\omega$$-$$\alpha_{A^{\prime}}$ and $L/\xi$$-$$\alpha_{A^{\prime}}$ plane respectively. From Fig.~\ref{Fig5}(a) we observe that the
probability for CAR at $Q$ is vanishingly small for all values of coupling constant and angle of incidence from channel $A^{\prime}$. 
The maximum CAR probability, that we can achieve in this case, is about $3\%$ which is significantly smaller in magnitude compared to 
that of the same for channel $A$. This reduction appears due to the momentum difference between $A$ and $A^{\prime}$.
This behavior is also almost independent of $\omega$. We present our result for $\omega=0.05-0.3$ eV which covers the experimentally 
achievable value~\cite{zhang2010crossover}. Although this phenomenon is true 
for a particular value of $L/\xi$, the result does not change by appreciable amount when we change the length of the superconducting
region. In Fig.~\ref{Fig5}(b), we show the corresponding behavior of CAR in $L/\xi$$-$$\alpha_{A^{\prime}}$ plane.
On the other hand, in contrast to CAR, CT is the dominating process for a wide range of $\omega$. This is evident from Fig.~\ref{Fig5}(c). 
Also note that, CT at normal incidence can dominate over the other
scattering processes even for weakly coupled surface states as shown in Fig.~\ref{Fig5}(c). However, such behavior is not 
entirely independent of the coupling constant. Apart from the weak coupling limit, a resonance can also be obtained at around 
$\omega=0.26$ eV. To reveal the behavior of CT at $Z$, as a function of the superconducting length, we present Fig.~\ref{Fig5}(d) 
where it is shown that the probability of CT at $Z$ manifests an oscillatory behavior. Although the amplitude of oscillation decreases as we
 \begin{figure}
 \begin{minipage}[t]{0.5\textwidth}
 \hspace{-.1cm}{ \includegraphics[width=.5\textwidth,height=4cm]{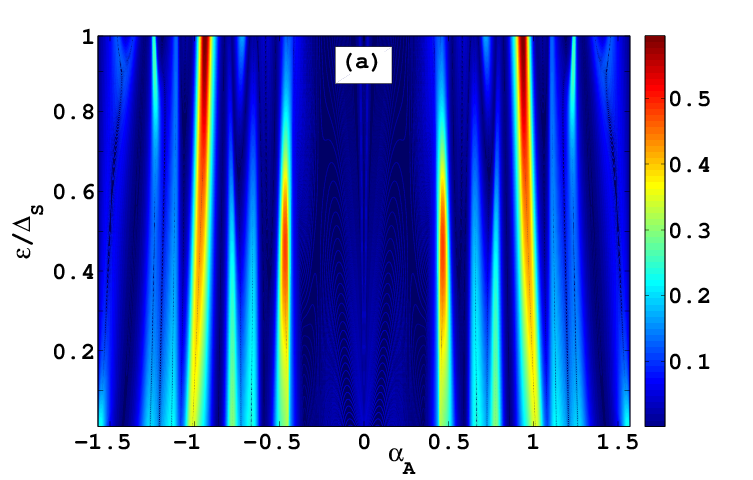}}
 \hspace{-.2cm}{ \includegraphics[width=.5\textwidth,height=4cm]{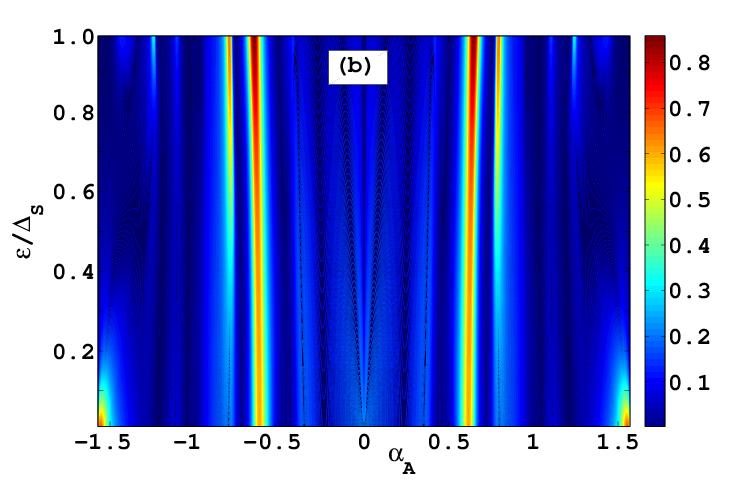}}
\end{minipage}
\caption{(Color online) The variation of CAR at $Q$ and CT probability at $Z$, in the $\epsilon/\Delta_{s}$$-$$\alpha_{A}$ plane, is illustrated in panel (a) and (b) respectively . 
The length of superconducting region is considered to be $L=0.5\xi$. The value of the other parameters are chosen to be $\omega=0.3$ eV and  $U=0.3$ eV.}
 \label{Fig6}
 \end{figure}
increase the length of the superconductor, even for normal incidence.
Here we present our result for $\omega=0.3$ eV for which we can achieve the maximum value of CT probability $\sim 0.6$. 
The oscillatory response of CAR and CT with the enhancement of the superconducting length is similar to the previous case 
(see Fig.~\ref{Fig3}). Nevertheless, the difference lies in the fact that in Fig.~\ref{Fig5}(d) the angular region, spanned by the angle 
of incidence for CT, is wider compared to that of depicted in Fig.~\ref{Fig3}(d). 

Similar to the case of incident channel at $A$, we can have CAR and CT probabilities at $J$ and $H$ also for the incident channel 
at $A^{\prime}$. Although they appear to be very small in magnitude for all values of $\omega$ and $L/\xi$. Hence, we do not show those 
results explicitly. The reason for vanishingly small CAR and CT probabilities for an incident electron from $A^{\prime}$ can be attributed to the 
small $x$-component of momentum in comparison to $A$.

Finally, we look into the behavior of CAR and CT probabilities with the variation of both incident electron (from point $A$) energy, below the 
subgapped regime, and incident angle ($\epsilon/\Delta_{s}$$-$$\alpha_A$ plane) as shown in Fig.~\ref{Fig6}. 
It is observed that CAR probability at $Q$ ($T^h_Q$) at a certain angle of incidence attains maximum value when energy 
of the incident electron becomes nearly equal to the proximity induced superconducting gap \ie $\eps\simeq \Delta_{S}$ (see Fig.~\ref{Fig6}(a)). 
On the oher hand, CT probability at $Z$ ($T^e_Z$) also exhibits similar behavior like CAR, as depicted in Fig.~\ref{Fig6}(b).
The behavior of CAR and CT for the $A^{\prime}$ channel is  also very similar to that of $A$ channel. 

\subsubsection{Case II: $\mu_R=-\mu_L$}

\begin{figure}[!thpb]
\centering
\includegraphics[height=4.0cm,width=0.8 \linewidth]{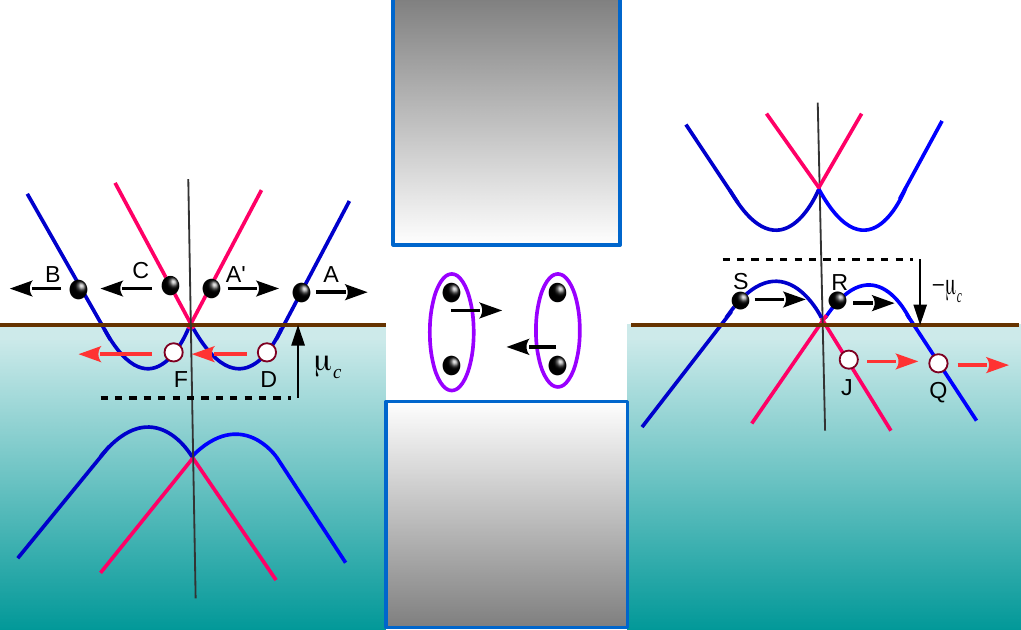}
\caption{(Color online) Schematic diagram of different scattering channels for our NSN hybrid structure when $\mu_R=-\mu_L$.
We adopt the same convention, mentioned in Fig.~\ref{Fig2}, to denote electrons and holes and their propagation direction.}
\label{Fig7}
\end{figure}

In this subsection, we discuss the scenario where the chemical potential in the $p$-type normal region is lifted to 
$\mu_{_R}=-\mu_{_L}=-\mu_c$. For this case, all possible scattering channels are shown in Fig.~\ref{Fig7}. Note that, now two 
CT channels belong to the same branch associated with a sign change in the $x$-component of momentum in one CT channel at $R$, \ie from 
$-k_{Z}^{x}\rightarrow k_{R}^{x}$.
The group velocity of the electron, along the $x$-direction, corresponding to the transmission channels at $R$ and $S$ 
would be positive as long as the slope $\partial E/\partial k_x>0$. This corresponds to the fact that group velocity along the $x$-direction 
can be positive even for negative $x$-component of momentum. Following the previous case, we also analyze here the behavior of 
CT and CAR probabilities, for an electron incoming from $A$. 
 
\begin{figure}[htb]
  \begin{minipage}[t]{0.5\textwidth}
  \hspace{-.1cm}{ \includegraphics[width=.5\textwidth,height=4cm]{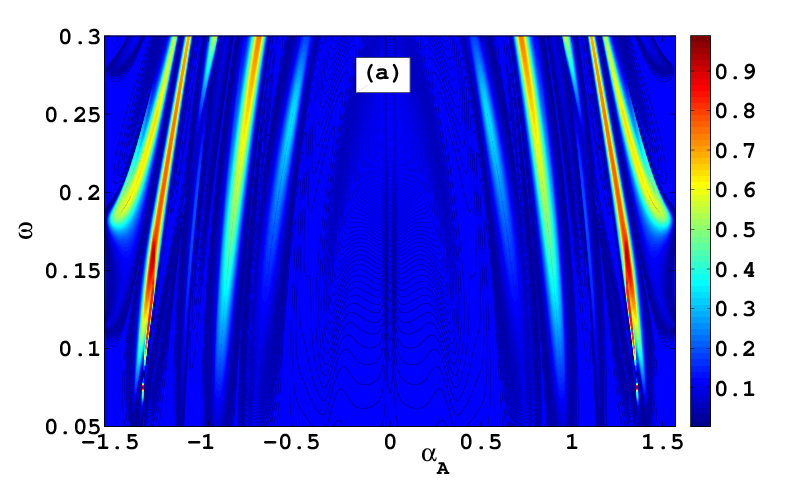}}
  \hspace{-.2cm}{ \includegraphics[width=.5\textwidth,height=4cm]{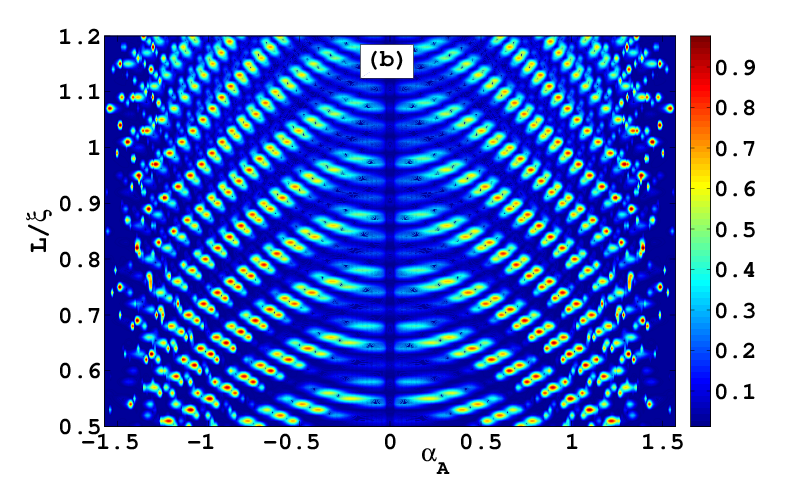}}
  \end{minipage}
  \begin{minipage}[t]{0.5\textwidth}
  \hspace{-.1cm}{ \includegraphics[width=.5\textwidth,height=4cm]{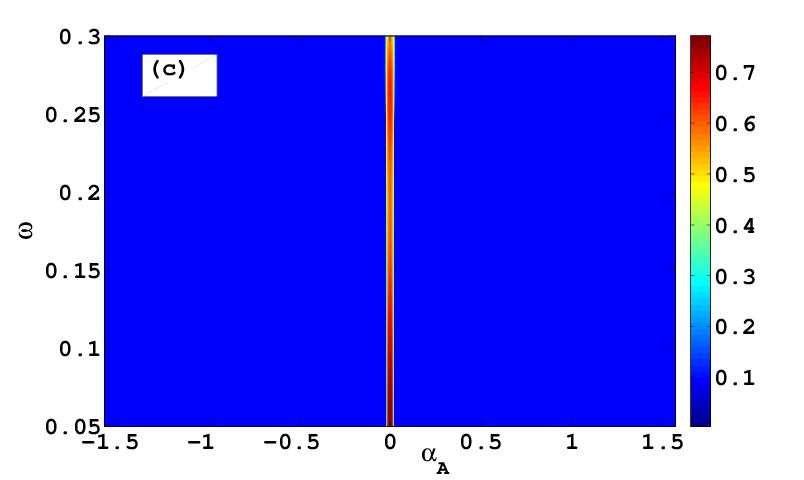}}
  \hspace{-.2cm}{ \includegraphics[width=.5\textwidth,height=4cm]{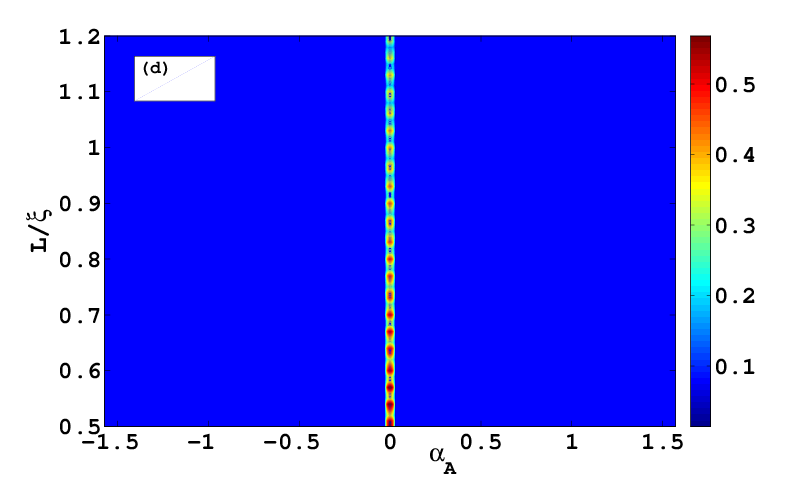}}
  \end{minipage}
\caption{(Color online) The behavior of CAR probability at $Q$ \ie $T_{Q}^{h}$ is shown in panel (a) and (b) in the plane of 
$(\omega,\alpha_A)$ and $(L/\xi,\alpha_A)$, respectively. Similarly, CT probability at $R$ \ie $T_R^{e}$ is demonstrated in 
the same parameter space in panel (c) and (d), respectively. The incident electron is considered to be from $A$. 
We choose the same values of the other parameters as mentioned in Fig.~\ref{Fig3}.}
 \label{Fig8}
 \end{figure}

In Fig.~\ref{Fig8}(a), we show the behavior of CAR probability at $Q$ as a function of the coupling strength ($\omega$) and angle of incidence 
($\alpha_A$) considering the same length of the \suc as mentioned in the previous cases. The most interesting feature is that CAR probability 
can be enhanced to more than $95\%$ in the $\omega$$-$$\alpha_{A}$ plane. This enhancement may be related to the matching of the $x$-component 
of the wave vector of the incident electron at $A$ with that of transmitted hole at $Q$. In particular, this feature appears for $\omega=0.16$ 
eV at a particular angle of incidence away from $\alpha_{A}=0$. On the other hand, CT at $R$ is allowed only through a very narrow angular 
region around $\alpha_A=0$ as shown in Fig.~\ref{Fig8}(c). The behavior of CAR and CT with the variation of the length of the superconducting 
region is illustrated in Figs.~\ref{Fig8}(b) and (d), respectively. The corresponding behavior of CAR probability at $Q$ preserves the rib-like 
pattern in the $L/\xi$$-$$\alpha_{A}$ plane as before (see Fig.~\ref{Fig8}(b)). Moreover, CT takes place with finite probability only around the 
normal incidence and decays with the enhancement of the length of the superconducting region as depicted in Fig.~\ref{Fig8}(d). Note that, CAR 
at $J$ is very small compared to $Q$. Also CT at $S$ exhibits similar behavior as of $H$ in the previous case. The contribution arising from 
incident electron at $A^{\prime}$ is also too small compared to that of $A$.
 
Therefore, by adjusting the chemical potential or the doping concentration in the right normal region from $\mu_{_R}=-1.5 \mu_c$ to $-\mu_c$, 
CAR probability can be maximally enhanced to $97\%$ even for finite angle of incidence (see Fig.~\ref{Fig8}(a)). This is the key result of our paper.
The striking feature is, unlike graphene, in thin TI hybrid junction a large CAR can be achieved even when both the normal regions are
sufficiently doped. One can achieve CAR with $100\%$ probability in graphene NSN junction under very special circumstances where 
the chemical potential in the $p$-doped region is chosen in such a way that CT channel falls exactly at the band touching point for 
which electron does not possess non-zero momentum, and hence only CAR mechanism is possible~\cite{cayssol2008crossed}. 
In massive Dirac material like silicene, chemical potential has to be adjusted at the bottom and top of the conduction (left region) 
and valence (right region) band, respectively,  so that energy band cannot support AR and CT due to the mass gap. As a result,
the incident electron can either be normally reflected or transmitted as a hole via the CAR process~\cite{PhysRevB.89.020504,paulquantum}.
Moreover, in silicene, this phenomena only occurs at normal incidence of electron \ie $\alpha_{A}=0$.
In this context, our system can be more advantageous in order to obtain large CAR probability without concomitant CT, but
with finite doping and oblique incidence. 

\begin{figure}[htb]
  \begin{minipage}[t]{0.5\textwidth}
  \hspace{-0.1cm}{ \includegraphics[width=.5\textwidth,height=4cm]{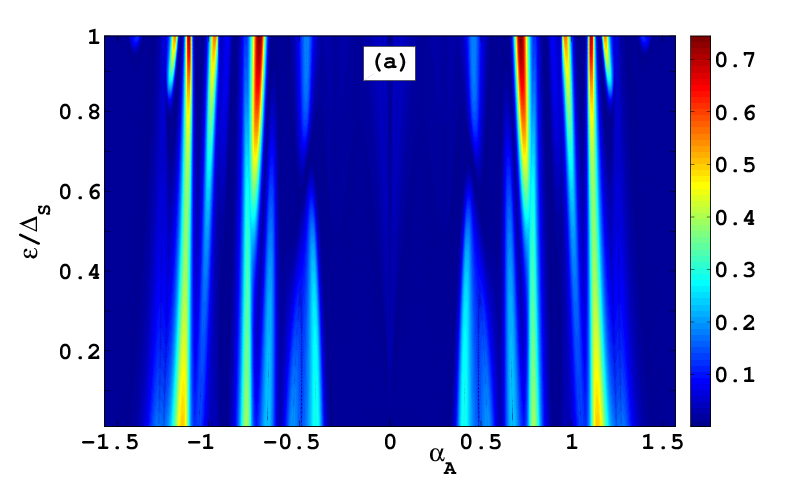}}
  \hspace{-0.2cm}{ \includegraphics[width=.5\textwidth,height=4cm]{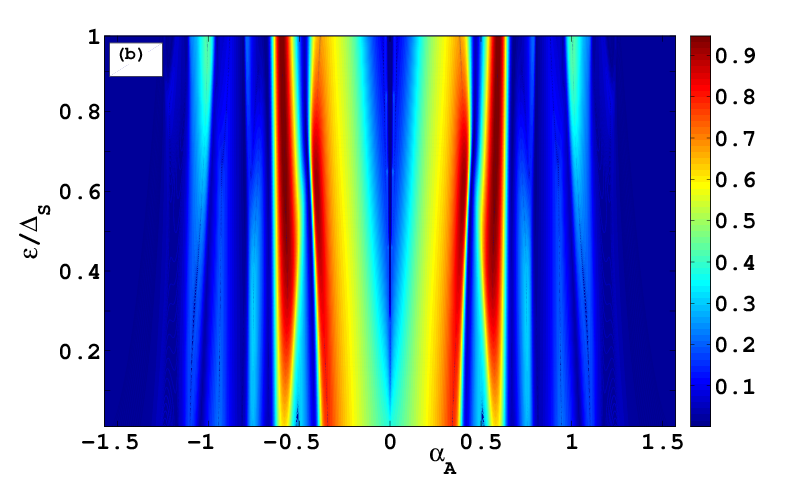}}
  \end{minipage}
  \caption{(Color online) The features of CAR at $Q$ and CT probability at $R$, in $\eps/\Delta_S$$-$$\alpha_{A}$ plane, are displayed in panel (a) and (b) respectively. 
The value of the other parameters are chosen to be the same as mentioned in Fig.~\ref{Fig6}.
}
\label{Fig9}
\end{figure}

In Fig.~\ref{Fig9}, we investigate the variation of CAR at $Q$ ($T^h_Q$) and CT probabilities at $R$ ($T^e_R$) in the $\eps/\Delta_{S}$$-$$\alpha_{A}$ parameter space.
Note that, CAR probability attains a maximum for $\eps\sim\Delta_{S}$ for a certain angle of incidence. This feature is depicted in Fig.~\ref{Fig9}(a).
Also this is very similar to the previous case (see Fig.~\ref{Fig6}(a)).
On the contrary, CT at $R$ can be quite high ($\sim 90\%$) for a wide range of incident electron energy $\eps$ at a particular angle of incidence.
This is shown in Fig.~\ref{Fig9}(b).  For the other incoming channel $A^{\prime}$, we have checked that both electron and hole transmissions are vanishingly small
in the same parameter regime.

\subsection{Conductance}
In this subsection, we investigate the angle-averaged normalized differential conductance for the two cases representing two different chemical 
potentials (doping concentration) in the right normal region. We employ extended BTK formalism~\cite{BTK,beenakker2006specular} to compute our conductance. 
The differential conductance corresponding to the elastic CT of electron at a particular angle of incidence ($\alpha_{j}$), where $j$ may be $A$ or $A'$, 
via transmission channel $\eta$ can be expressed as
\begin{equation}
G_{\rm CT}=\frac{e^2}{h}\sum_{j}\sum_{\eta}|t_{\eta}^{e}(\alpha_j)|^2.
\end{equation}
The transmission channel $\eta$ can be $\{Z,H\}$ or $\{R,S\}$ depending on the doping level either at $\mu_{_R}=-1.5\mu_{_L}$ 
or at $\mu_{_R}=-\mu_{_L}$, respectively.
\begin{figure}[htb]
\begin{minipage}[t]{0.5\textwidth}
  \hspace{-.6cm}{ \includegraphics[width=.55\textwidth,height=4.8cm]{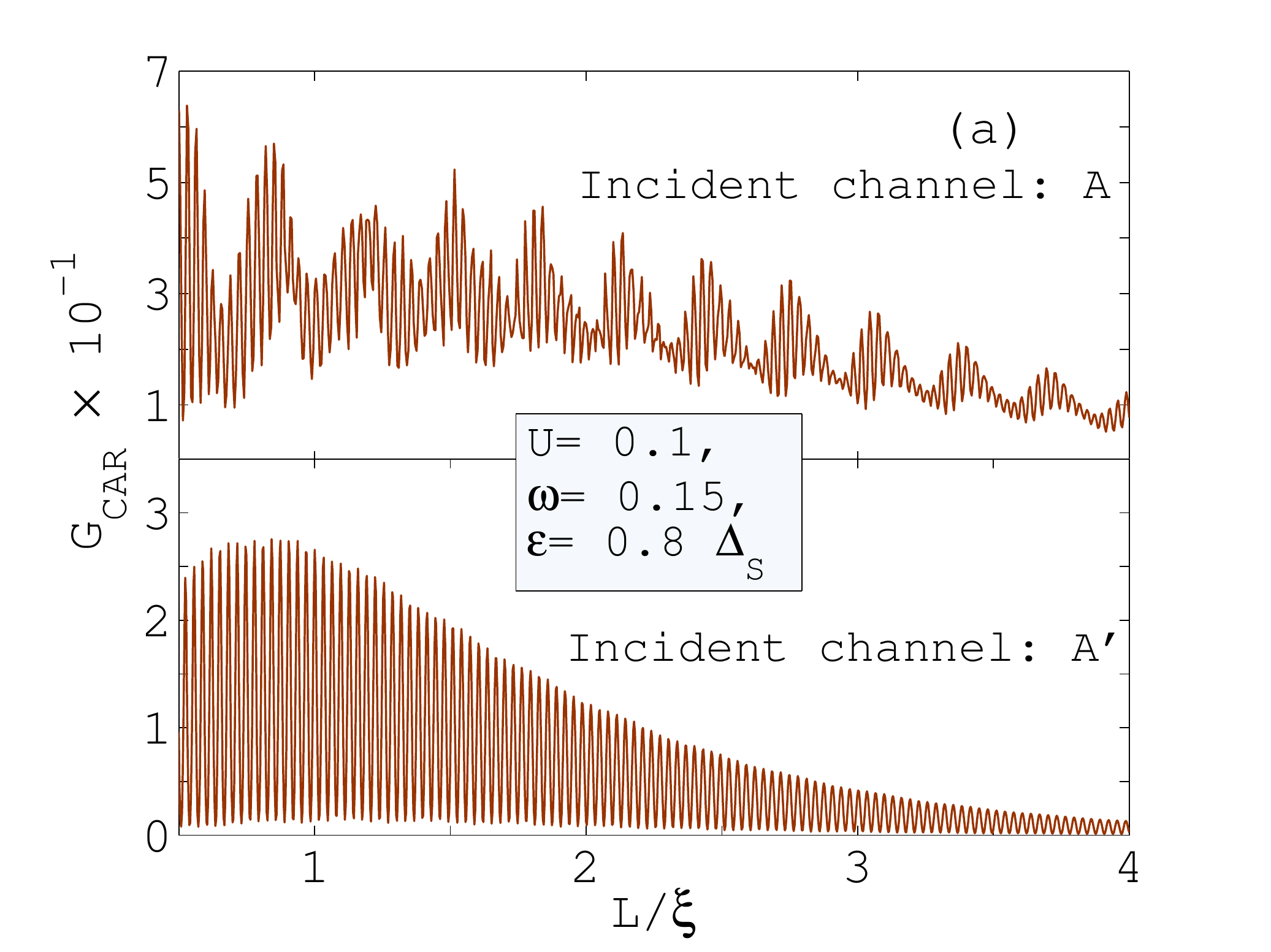}}
  \hspace{-.7cm}{ \includegraphics[width=.55\textwidth,height=4.8cm]{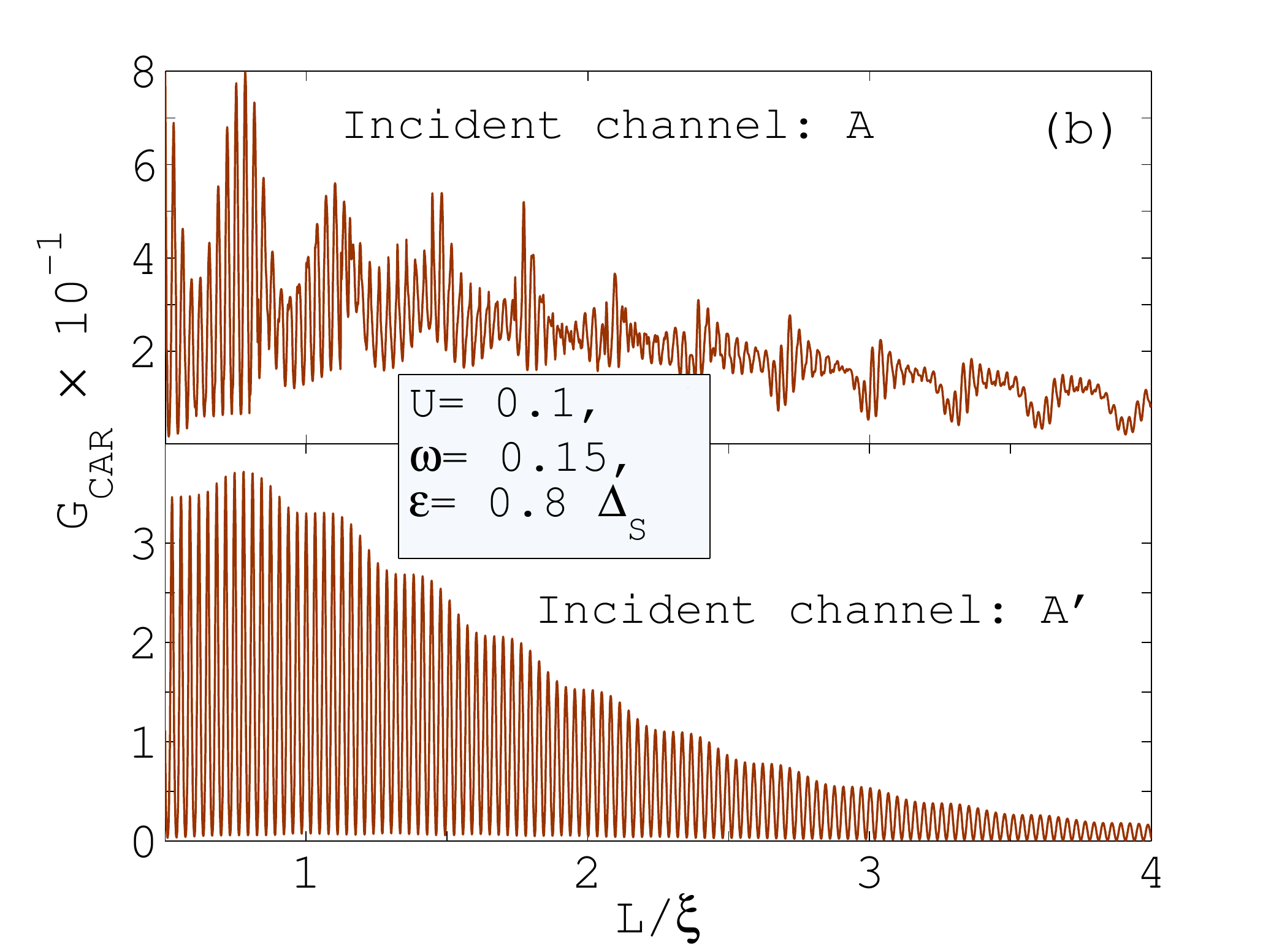}}
\end{minipage}
 \caption{(Color online) The behavior of angle averaged differential CAR conductance, in units of $G_{0}$, is shown as a function of the length of the superconductor. 
Here, $\omega$ and $U$ are in units of $\rm eV$. Panel (a) and (b) correspond to the case $\mu_{R}=-1.5\mu_{L}$ and $\mu_R=-\mu_L$
respectively.
}
 \label{Fig10}
 \end{figure}
After averaging over the angle of incidence the conductance reduces to
\begin{equation}\label{cond}
\frac{G_{\rm CT}}{G_0}=\sum_{j}\int_{-\pi/2}^{\pi/2}\sum_{\eta}|t_{\eta}^{e}(\alpha_j)|^2\cos \alpha_{j} d\alpha_{j} \ ,
\end{equation}
where $G_0=(e^2/h)N$, $N=kW/\pi$ being the number of transverse modes and $W$ is the width of the thin TI. Similar expression 
can be used for the CAR conductance as well
\begin{equation}\label{cond1}
 \frac{G_{\rm CAR}}{G_0}=\sum_{j}\int_{-\pi/2}^{\pi/2}\sum_{\bar{\eta}}|t_{\bar{\eta}}^{h}(\alpha_j)|^2\cos \alpha_j d\alpha_j\ .
\end{equation}
where $\bar{\eta}$ can be $\{Q, J\}$. Finally we compute the total two terminal differential conductance by using the relation
\begin{equation}
\frac{\partial I}{\partial V}=G=\frac{G_{\rm CT}-G_{\rm CAR}}{G_0} \ .
\end{equation}

In Fig.~\ref{Fig10}, we illustrate the behavior of angle averaged CAR conductance for an incoming electron from $A$ (upper row of Fig.~\ref{Fig10}(a)-(b)) 
and $A^{\prime}$ (bottom row of Fig.~\ref{Fig10}(a)-(b)) with the variation of the length of the superconductor. Here we also investigate how the
thickness induced coupling parameter $\omega$ and the doping level $\mu_{R}$ in the right normal region affect the CAR conductance.
We find that, when $\mu_{R}=-1.5\mu_{L}$, CAR conductance exhibits an oscillatory behavior as we vary $L/\xi$ (see Fig.~\ref{Fig10}(a)). 
This result is consistent with the previous oscillatory nature of CAR probability. Such oscillatory nature of CAR conductance can be a direct 
manifestation of closely spaced Andreev bound state levels inside the superconducting region. Moreover, change of incoming channel seems to 
induce an additional oscillation envelope due to the coupling between the two TI surfaces, as
depicted in the upper row of Fig.~\ref{Fig10}(a). Such oscillation in CAR conductance can be observed when the electron is incident via channel $A$. 
However, this additional oscillation is not visible for the incident electron at $A'$ (see lower row of Fig.~\ref{Fig10}(a)). 
For $\mu_R=-\mu_L$ the behavior as well as the magnitude of the oscillation appears to be similar as $\mu_{R}=-1.5\mu_{L}$ case (see Fig.~\ref{Fig10}(b)).
Note that, similar oscillatory behavior of the angle averaged CT conductance also exists. 

\begin{figure}[htb]
\begin{minipage}[t]{0.5\textwidth}
  \hspace{-.4cm}{ \includegraphics[width=.5\textwidth,height=4cm]{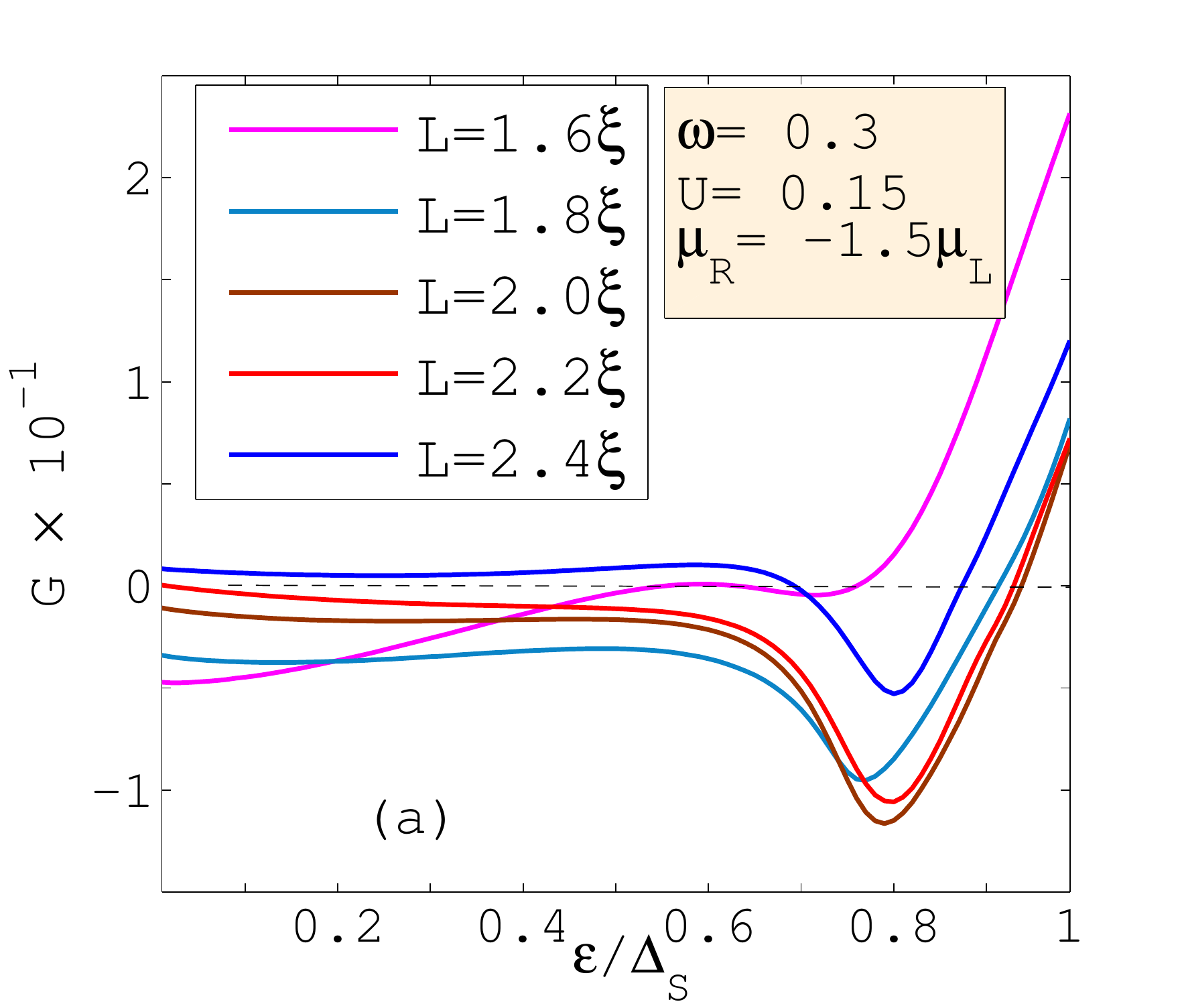}}
  \hspace{-.5cm}{ \includegraphics[width=.5\textwidth,height=4cm]{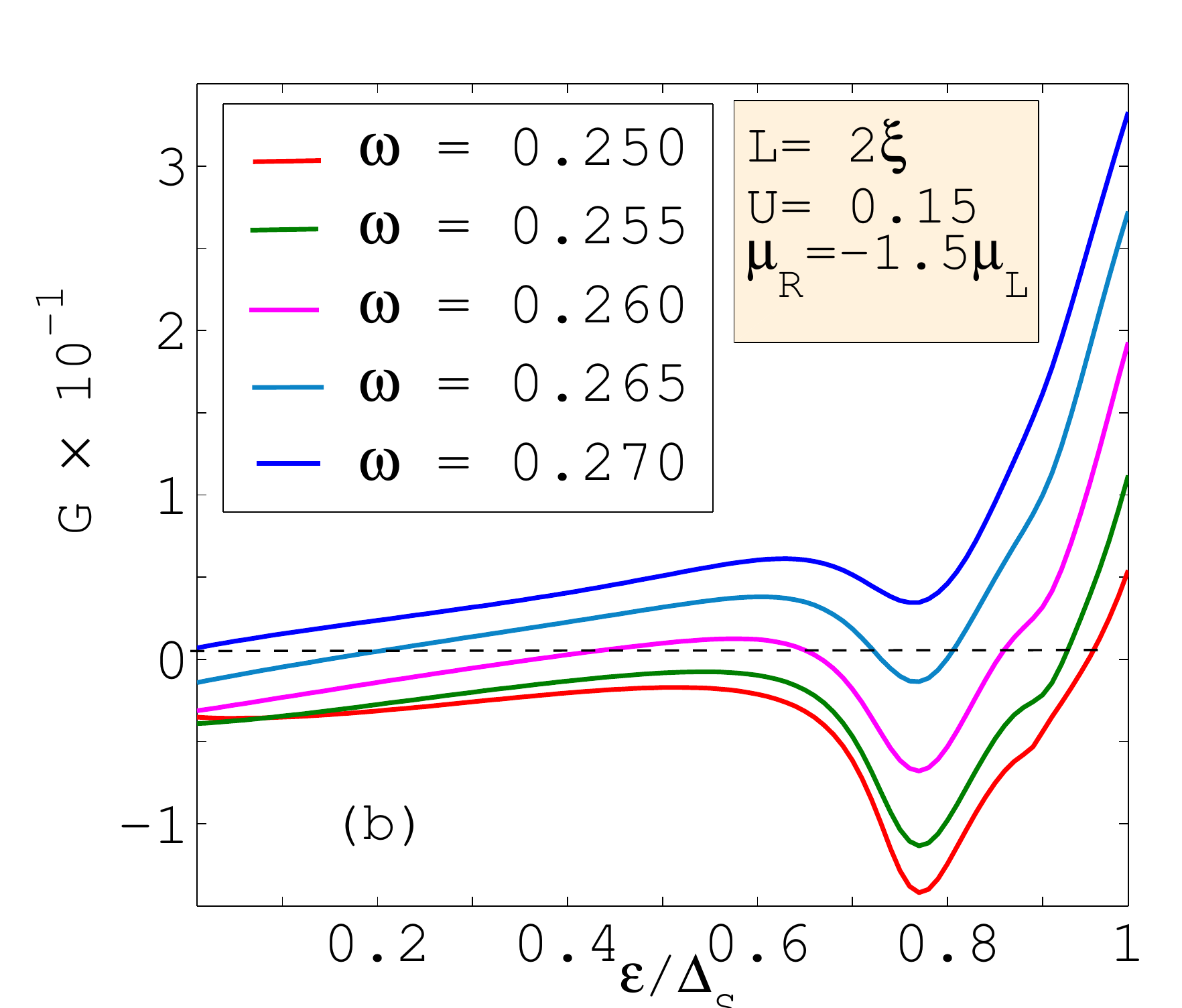}}
\end{minipage}
\begin{minipage}[t]{0.5\textwidth}
  \hspace{-.4cm}{ \includegraphics[width=.5\textwidth,height=4cm]{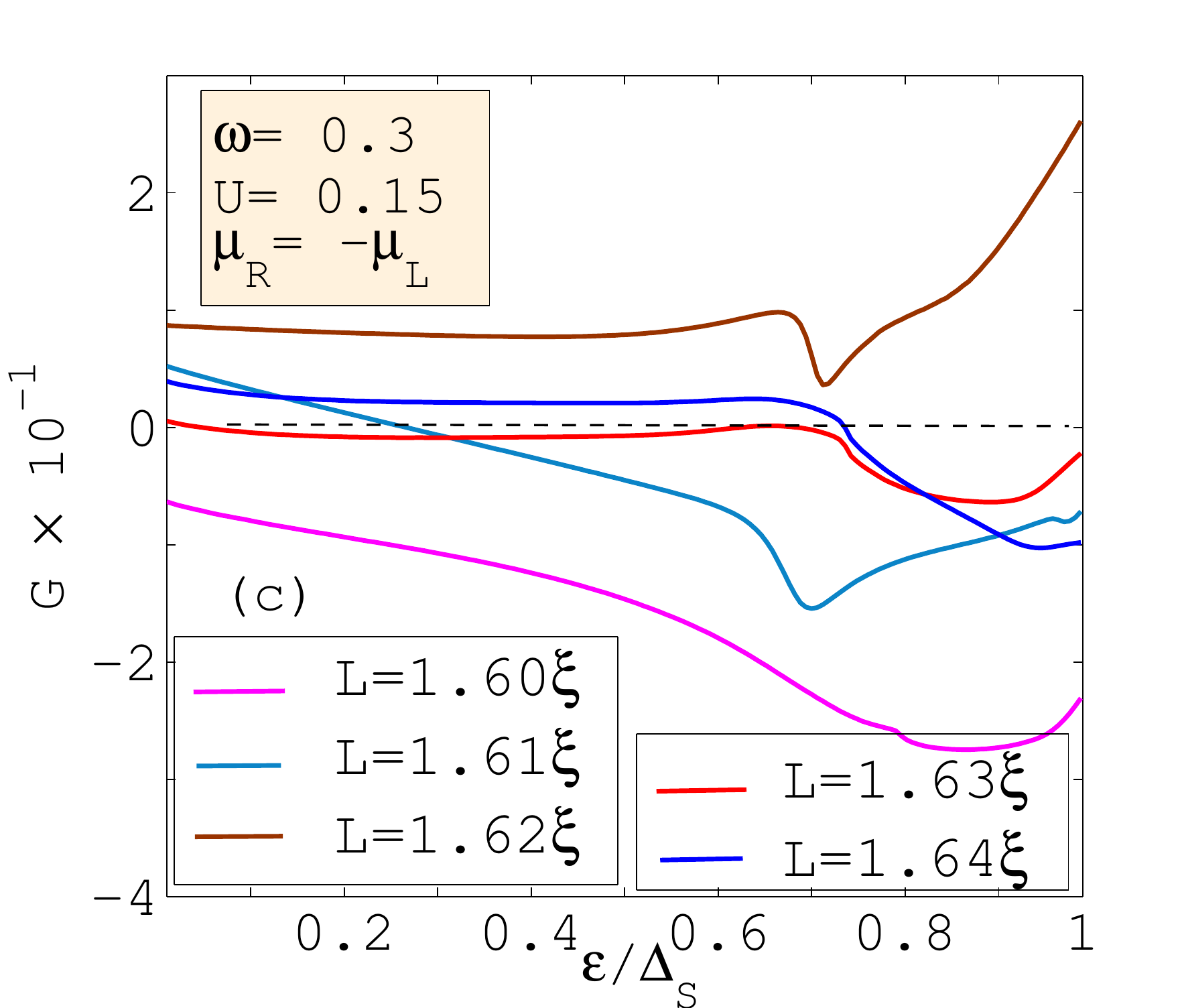}}
  \hspace{-.5cm}{ \includegraphics[width=.5\textwidth,height=4cm]{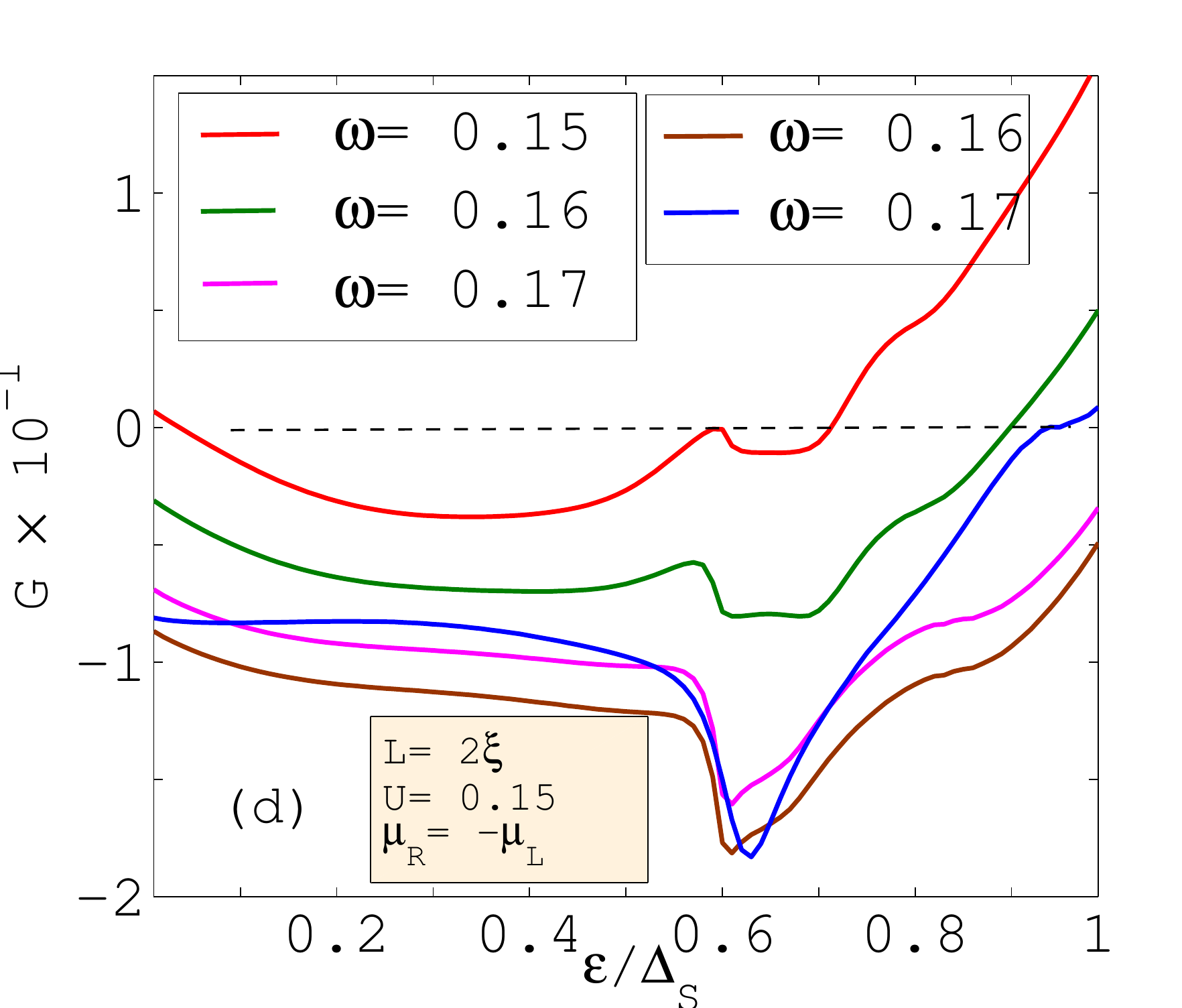}}
\end{minipage}
 \caption{(Color online) The behavior of angle averaged differential conductance $G=G_{CT}-G_{CAR}$, in units of $G_0$,
 is demonstrated as a function of $\epsilon/\Delta_{S}$ in the subgapped regime. Here, $\omega$ and $U$ are in the units of $\rm eV$.}
 \label{Fig11}
 \end{figure}

In Fig.~\ref{Fig11}, we show the behavior of angle averaged differential conductance $G$ (in unit of $G_{0}$) as a function of the incident electron 
energy $\epsilon$ in the subgapped regime ($\epsilon\leq \Delta_{s}$) incorporating contributions arising from both the incident channels $A$ 
and $A^{\prime}$. Here, the doping level in the right normal region is $\mu_R=-1.5\mu_L$. From Figs.~\ref{Fig11}(a)-(b), it is apparent that 
CAR can dominate over CT within the regime $\eps \le 0.8 \Delta_{s}$. However, this phenomenon is very sensitive to the coupling $\omega$ as 
well as the length of the superconducting region $L$. In fact, a smooth variation of $\omega$ can reduce the CT conductance resulting in higher
CAR contribution which can be observed in Fig.~\ref{Fig11}(b). The competition between these two scattering processes give rise to positive or 
negative values of relative $G$. The results are of same nature for $\mu_R=-\mu_L$ case which is shown in Figs.~\ref{Fig11}(c)-(d). It can be 
seen that the coupling between the two TI surfaces enhances CAR conductivity by a larger amount compared to that of $\mu_R=-1.5\mu_L$. Therefore, 
in our NSN hybrid structure, CAR conductance can dominate over the CT conductance, below the subgapped regime, under suitable circumstances.

 \subsection{Shot noise}

This subsection is devoted to the analysis of zero-frequency shot noise cross correlation for our hybrid NSN model following 
Refs.[\onlinecite{blanter2000shot, PhysRevB.53.16390}]. 
The current-current correlation function between the two leads, labeled by $i$ and $j$, is given by
 \begin{equation}\label{corr}
  S_{ij}(t-t')=\la\Delta\hat{I}_{i}(t)\Delta\hat{I}_{j}(t')+\Delta\hat{I}_{j}(t')\Delta\hat{I}_{i}(t)\ra\ ,
 \end{equation}
where the current fluctuation operator is defined as
\begin{equation}
 \Delta\hat{I}_{i}(t)=\hat{I}_{i}(t)-\la\hat{I}_{i}(t)\ra \ .
\end{equation}
The correlation function defined in Eq.(\ref{corr}) can be transformed in Fourier space as
\begin{equation}
 S_{ij}(\Omega)\delta(\Omega+\Omega')=\frac{1}{2\pi}\la\Delta\hat{I}_{i}(\Omega)\Delta\hat{I}_{j}(\Omega')+\Delta\hat{I}_{j}(\Omega')
 \Delta\hat{I}_{i}(\Omega)\ra\ ,
\end{equation}
with
\begin{equation}
 \Delta\hat{I}_{i}(\Omega)=\hat{I}_{i}(\Omega)-\la\hat{I}_{i}(\Omega)\ra\ .
\end{equation}
Now using $\int dt~ e^{(\eps-\eps')t/\hbar}=2\pi\hbar\delta(\eps-\eps')$, the zero frequency shot noise cross-correlation between 
the two leads, $i$ and $j$, in terms of scattering amplitudes can be generalized in presence of an external bias as~\cite{PhysRevB.53.16390}
\begin{eqnarray}
 S_{ij}(\eps)&=&\frac{2e^2}{h}\sum_{k,l\in N,S;\alpha,\beta,\gamma,\delta\in e,h}sgn(\alpha)sgn(\beta)\nonumber\\&& A_{k\gamma,l\delta}(i\alpha,\eps)
 A_{l\delta,k\gamma}(j\beta,\eps)f_{k\gamma}(\eps)[1-f_{l,\delta}(\eps)]\ , \nonumber\\
\label{sij}
 \end{eqnarray}
where $A_{k\gamma,l\delta}(i\alpha,\eps)=\delta_{ik}\delta_{il}\delta_{\alpha\gamma}\delta_{\alpha\delta}-s_{ik}^{\alpha\gamma^{\dagger}}(\eps)
s_{il}^{\alpha\delta}(\eps)$. Here, $s_{ik}^{\alpha\gamma}$ denotes the scattering amplitude for a $\gamma$ type particle incident from lead $k$ 
being scattered to lead $i$ as a particle type $\alpha$ ($\alpha,\gamma\in e, h$) where $e$ ($h$) stands for electron (hole). Also $sgn(\alpha)$ ($sgn(\beta)$)
can be $+~(-)$ corresponding to $e$ ($h$). As the scattering amplitudes are function of angle of incidence, we perform the angle average of shot noise 
cross-correlation incorporating the contributions arising from both the incoming channels of the incident electron ($A$ and $A'$),
\begin{equation}\label{shot}
 S_{ij}(\eps)=\int_{-\pi/2}^{\pi/2}S_{ij}(\alpha_A,\eps)d\alpha_{A}+\int_{-\pi/2}^{\pi/2}S_{ij}(\alpha_{A'},\eps)d\alpha_{A'} \ .
\end{equation}
A simplified analytical expression of the zero frequency shot noise cross-correlation that we use for our numerical computation is given in Appendix~\ref{App2}.
\begin{figure}[htb]
{\includegraphics[width=.5\textwidth,height=8.5cm]{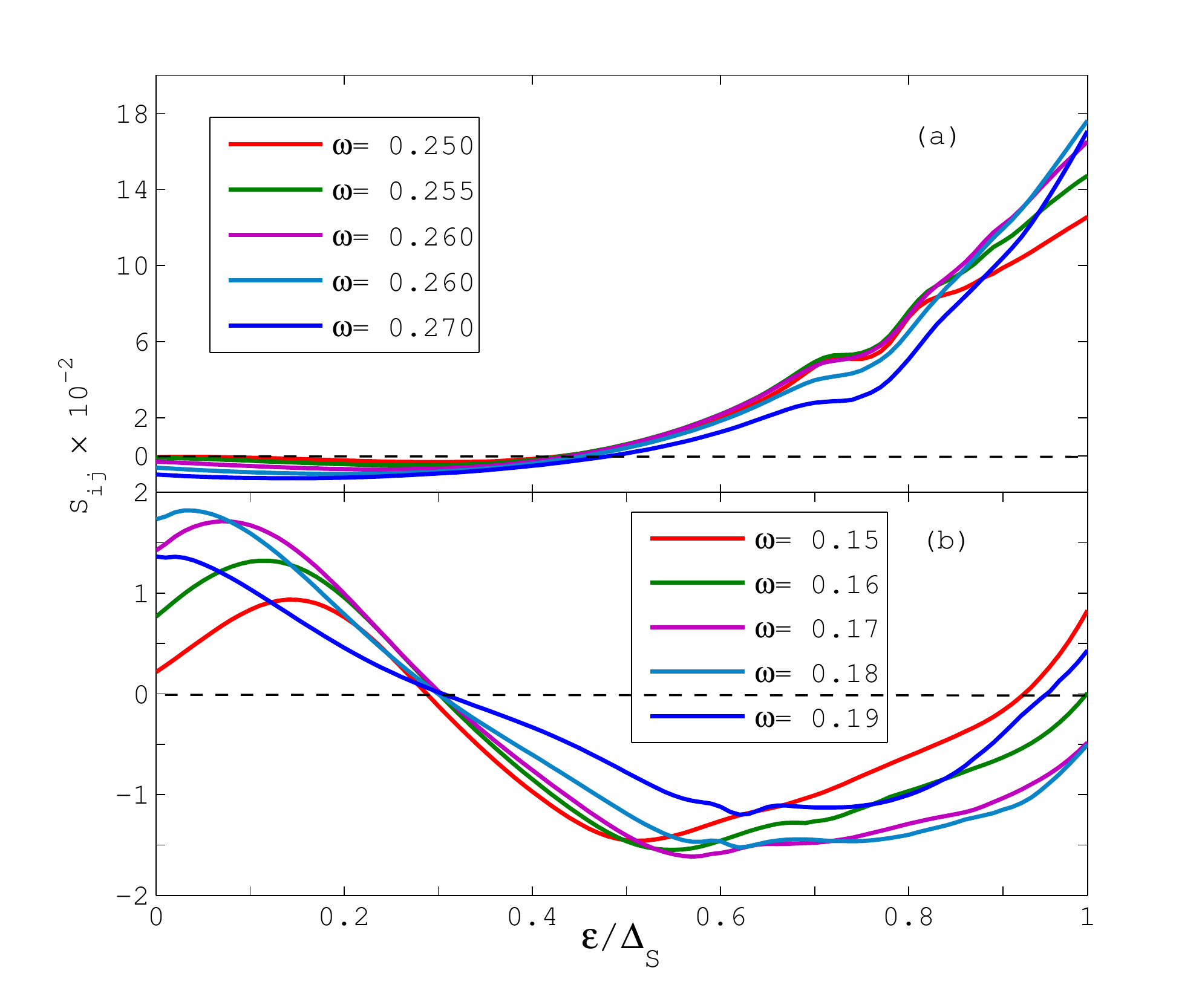}}
\caption{The behavior of shot noise cross correlation $S_{ij}$ (in units of $e^2/h$) is illustrated as a function of $\epsilon/\Delta_{S}$ 
in the subgapped regime. Here panel (a) and (b) correspond to the case (a) $\mu_R=-1.5\mu_L$ and (b) $\mu_R=-\mu_L$ respectively
and $\omega$ is in the unit of $\rm eV$.
The value of the other parameters are $U=0.1$ eV and $L=2\xi$.}
 \label{Fig12}
 \end{figure}
 
In Fig.~\ref{Fig12} we show the features of the shot noise cross-correlation $S_{ij}$ with the variation of the incoming electron energy 
$\epsilon$ below the subgapped regime ($\epsilon \leq \Delta_{S}$). 
We observe that shot noise changes sign from negative to positive depending on the incoming electron energy as well as the coupling $\omega$
between the two TI surfaces. 
When the chemical potential in the right normal region is set to the value $\mu_R=-1.5 \mu_L$ we observe that 
the transition of $S_{ij}$ from negative to positive is monotonic. After a critical value of the incoming electron energy (below $\Delta_{S}$), 
$S_{ij}$ crosses over to the positive value for all values of the coupling strength $\omega$. This is shown is Fig.~\ref{Fig12}(a).

On the other hand, for $\mu_R=-\mu_L$ the crossover behavior of $S_{ij}$, from positive to negative, appears to be non-monotonic. 
When the incoming electron energy is well below the proximity induced superconducting gap $\Delta_S$,  $S_{ij}$ is positive. In this
regime, CAR is the dominating scattering process. However,  $S_{ij}$ changes it's sign to negative at $\epsilon=0.3 \Delta_{S}$ and
CT becomes dominating over CAR. It can again be tuned to positive value for $\epsilon\sim\Delta_{S}$ and by tuning the coupling $\omega$ (see Fig.~\ref{Fig12}(b)). 
This happens due to the large contribution of CAR process in this parameter regime (see Figs.~\ref{Fig8}(a)-(b)).
Nevertheless, it is well-known that shot noise cross-correlation between two leads is always negative for fermions~\cite{blanter2000shot}. 
Note that, shot noise cross-correlation has been verified to be positive for $s$-wave superconductor in other systems under suitable 
parameter regime~\cite{thierrymartin1,thierrymartin2,WeiChandrasekhar,AndyDas}, which can also be the possible signature 
of spin entangle states~\cite{recher,samuelsonbuttiker,LesovikMatin,yeyati,PhysRevB.78.235403}. 
This type of cross-over phenomena of $S_{ij}$, from positive to negative, has also been reported in the context of transition from 
Majorana to Andreev bound states in a Rashba nanowire hybrid junction~\cite{haim2015signatures}. 
Whereas, in our case the transition of $S_{ij}$ from positive to negative or vice-versa is completely associated with the relative strength 
of the CAR and CT scattering process. The competition between these two processes leaves a signature on the shot noise cross-correlation 
being consistent with our earlier results of scattering amplitudes. Depending on the various parameters like, doping level $\mu_{R}$ in the right
normal region, coupling $\omega$ between the two TI surface states, length $L$ of the proximity induced superconducting region and 
incoming electron energy $\epsilon$, we can have both the positive and negative shot noise cross-correlation accordingly. 
 
 \section{Summary and Conclusions}\label{sec4}
To summarize, in this article, we explore the subgapped transport and shot noise properties of a NSN hybrid junction made of thin TI. 
We have considered normal region in the left and right side of the proximity induced superconducting region respectively as $n$-type and $p$-type, designing a 
$nSp$ type NSN junction. The coupling between the top and bottom surface states of thin 2D TI opens up a possibility to enhance the CAR probability 
in Dirac material. We observe that CAR probability can be achieved up to $97\%$ by tuning the gate voltage. In our case, the CAR is of specular type.
In contrast to graphene, $nSp$-type heterostructure of thin TI can be a suitable system for obtaining higher CAR probability even for an arbitrary doping concentration 
in the $p$-type region and finite angle of incidence. We consider two cases describing two different positions of the chemical potential (doping concentration) in the 
right normal region which are $\mu_R=-1.5\mu_L$ and $\mu_R=-\mu_L$. The maximum value of CAR probability is achieved to be around $92\%$ and $97\%$, respectively 
in the two cases while the rest is normal reflection probability. Moreover, the behavior of CAR conductance is rapidly oscillating with the length of the superconductor. 
Coupling between the two TI surfaces not only enhances CAR conductivity but also induces additional oscillation associated with much lesser frequency forming 
an envelope over the CAR conductance oscillation with the length of the superconductor. Depending on the choice of suitable parameter regime, 
CAR conductance can dominate over CT conductance or vice-versa. Furthermore, we also investigate the shot noise cross correlation and show that 
noise correlation may exhibit monotonic or non-monotonic behavior depending on the doping concentration of the $p$-type region. Below the subgapped regime, 
it monotonically decreases with the increase of incoming electron energy for $\mu_R=-1.5\mu_L$. On the other hand, the behavior becomes non-monotonic as we 
change the doping level to $\mu_R=-\mu_L$.
Shot noise can even change sign from negative to positive or vice-versa depending on the parameter values. In our case, this sign changing feature is associated 
with the relative strength of the CAR and CT probability occurring at the right interface of our NSN hybrid junction. The positive sign of shot noise cross-correlation 
can be a possible signature of entangled states~\cite{recher,samuelsonbuttiker,LesovikMatin,yeyati} in Dirac systems. 

As far as practical realization of our NSN hybrid structure is concerned, superconducting correlation can be induced inside thin film of 
Bi$_2$Se$_3$ via the proximity effect. This has been recently demonstrated in Refs.~[\onlinecite{wang2012coexistence,PhysRevX.4.041022}] 
by using $\rm NbSe_2$ superconductor with gap $\Delta_S \sim 1.5 $~meV. The coupling strength lies in between $\omega\approx 0.05-0.25$~eV 
corresponding to the range of thickness $5-2$ nm~\cite{zhang2010crossover}, which has been considered in our analysis. As the strength of 
the coupling between the two TI surface states cannot be tuned externally, one can vary the gate voltage $U$ to split the bands in order 
to obtain maximum non-local conductance for a fixed $\omega$. The present model may also be used as a beam splitter device to obtain Cooper 
pair splitting with efficiency higher than other systems~\cite{WeiChandrasekhar,AndyDas}.

Finally, it is important to mention that the enhancement of CAR in our thin TI hydrid structure is not only the monopoly of strict parameter regime ($\epsilon=\Delta$). 
The CAR probability, higher than $95\%$, can be acheived even if the exciation energy is less than the superconducting gap ($\epsilon<\Delta$) and also for a 
wide range of angle of incidence by adjusting the gate voltage appropriately. This is one of the main advantages of any layered system that one 
can tune the gate voltage to modulate transport properties (CAR process in our case) of the system. One can also be curious about the behavior of 
superconducting order parameter under the combined effects of the gate voltage and the coupling between the two surfaces. In order to make this confusion clear, 
we have also checked our main results by solving the standard self consistency of the gap equation for BCS theory. We have noticed that the gap parameter varies very slowly 
with the gate voltage and the hybridization, so the assumption of constant gap parameter is well justified for our study. Also our calculation is valid for zero temperature where
the self-consistent variation of the superconducting gap parameter with temperature is unimportant. 

 \begin{appendix}
  \section{Low energy effective Hamiltonian for thin topological insulator}
 \label{App}
We consider the two surfaces of a thin TI lying in $x$-$y$ plane. The thickness of the film is $d$ along $\hat{z}$-direction.
The electrons are confined along the $\hat{z}$-direction but free to move in $x$-$y$ plane. So, we can assume
${\bf p}=\{p_x,p_y\}$ as good quantum number. The single Dirac cone in each surface can be modeled by Rashba
Hamiltonian as
\begin{equation}
 h_{t(b)}=\pm v_{F}(\vec{\sigma}\times \vec{p})_{z}
\end{equation}
where $+(-)$ corresponds to top (bottom) surface, $v_{F}$ is the Fermi velocity and $\vec{p}=\{p_x,p_y\}$
is $2$D momentum. The coupling parameter $\omega$ between the top and bottom surfaces can be included in the 
total Hamiltonian as
\begin{equation}
 H=\left[\begin{array}[c]{c c}
          h_{t} &\omega\sigma_0\\ \omega\sigma_0& h_{b}
         \end{array}\right],
\end{equation}
where $\sigma_0$ is a $2\times2$ identity matrix. The above equation can be rewritten as
{\scriptsize
\begin{equation}
 H=\left[\begin{array}[c]{c c c c}
          0& v_{F}(p_y+ip_x)&\omega & 0\\v_{F}(p_y-ip_x)&0&0&\omega\\
          \omega&0&0&-v_{F}(p_y+ip_x)\\
          0&\omega&-v_{F}(p_y-ip_x)&0
         \end{array}\right].
\end{equation}
}
Now, we consider that the top surface is connected to the potential $V_{t}=U$ and the bottom surface is at $V_{b}=-U$.
Such arrangement introduces a potential difference $2U$ between the two surfaces. This potential difference can be inserted into 
the above Hamiltonian as
{\scriptsize
 \begin{equation}
 H=\left[\begin{array}[c]{c c c c}
          U& v_{F}(p_y+ip_x)&\omega & 0\\v_{F}(p_y-ip_x) & U & 0 &\omega\\
          \omega & 0 &-U & -v_{F}(p_y+ip_x)\\
          0 & \omega &-v_{F}(p_y-ip_x) &-U
         \end{array}\right].
\end{equation}
}
The above Hamiltonian can be further written in a compact form as
\begin{eqnarray} 
H&=&\left[\begin{array}[c]{c c}
          h(p)+U\sigma_{0} &\omega\sigma_{0}\\ \omega\sigma_{0}& -h(p)-U\sigma_{0}
         \end{array}\right]\nonumber\\
      &=&\hat{\tau}_{z}\otimes\hat{h}(p)+\hat{\tau}_x\otimes\omega\hat{\sigma}_0+U\hat{\tau}_z\otimes\hat{\sigma}_0.
\end{eqnarray}

This Hamiltonian well describes our set-up.

 \section{The wave functions in three different regions of our NSN hybrid structure}
 \label{App1}
The basic ingredients for solving the scattering problem are the wave functions in three different regions. In our case, the chemical 
potential in the left-normal region is adjusted at $\mu_{L}=\mu_c$ for which the wave functions are  already evaluated in 
Ref.~[\onlinecite{PhysRevB.93.195404}]. Following that we write the wave function in the left region as
\begin{equation}
 \Psi_{L}=\psi_{in,j}^{e}+\sum_{\iota}\psi_{r,\iota}^{e}+\sum_{\eta}\psi_{r,\eta}^{h}\ ,
\label{A1}
\end{equation}
where the first term in Eq.(\ref{A1}) is the wave function of the incident electron at $j=A$ or $A'$ 
which can be written as
\begin{equation}
 \psi^{e}_{in,j}=A_{j}^{e}e^{ik_{j}^{x}x}\left[\begin{array}[c]{c}
               1\\
               -ia_{j}^{e}e^{i\alpha_j}\\
               b_j^{e}\\
               -ic_j^{e}e^{i\alpha_j}\\0\\0\\0\\0\\
              \end{array}\right].
\end{equation}
The second term stands for the normally reflected electron from $\{\iota=B,C\}$ and reads as
\begin{equation}
 \psi^{e}_{r,\iota}=A_{\iota}^{e}r_{\iota}^{j}e^{-ik_{\iota}^{x}x}\left[\begin{array}[c]{c}
              1\\
               ia_{\iota}^{e}e^{-i\alpha_{\iota}}\\
               b_{\iota}^{e}\\
               ic_{\iota}^{e}e^{-i\alpha_{\iota}}\\0\\0\\0\\0\\
              \end{array}\right].
\end{equation}
Finally, the reflected hole state \ie Andreev reflected part corresponding to the last term in Eq.(\ref{A1}) can be expressed as
 \begin{equation}
 \psi_{r,\eta}^{h}=A_{\eta}^{h}e^{ik_{\eta}^{x}x}\left[\begin{array}[c]{c}
               \\0\\0\\0\\0\\
               1\\
               -ia_{\eta}^{h}e^{i\alpha_\eta}\\
               b_\eta^{h}\\
               -ic_\eta^{h}e^{i\alpha_\eta}\\
              \end{array}\right]
\end{equation}
for $\eta=D$. The reflected hole state at $F$ can be obtained by substituting $\eta=F$ with $\alpha_D\rightarrow(\pi-\alpha_F)$
and $k_{D}^{x}\rightarrow -k_{F}^{x}$ where $\alpha_i=\tan^{-1}(k_y/k_i^{x})$, with $i=j,\iota,\eta$, denote the angle of 
incidence for $j$ and reflection for $\iota$ and $\eta$. Here, $r_{\iota}^{j}$,$r_{\eta}^{j}$, are the reflection amplitudes of 
electron and hole, respectively. Also the $x$-component of wave vector is $k_{i}^{x}=\sqrt{(k_i)^2-k_y^2}$. The factor $A_{i}^{e(h)}$ 
is included to satisfy the conservation of probability current density during scattering mechanism, 
which is given by
\begin{equation}
 A_{i}^{e(h)}=1/\sqrt{(a_i^{e(h)}-b_{i}^{e(h)}c_i^{e(h)})\cos{\alpha_i}}\ .
\end{equation}
Note that, there is a critical angle of incidence beyond which no reflection occurs in the left region. 
To evaluate this critical 
angle, we use the fact that $k_{i}^{x}$ as well as $\alpha_i$ becomes imaginary beyond critical angle \ie $\alpha_{_C}=-i\chi$ 
with $\chi=\tanh^{-1}\Big[k_y/\sqrt{k_y^2-k_C^2}\Big]$. The angle of reflection becomes imaginary for reflection channel $C$, when
$k_y\ge k_C$ where $|k_{C}|=\Big[\sqrt{(\eps+\mu_N)^2-\omega^2}-U]/(\hbar v_F)\Big]$. 
At this condition, the critical angle of incidence is given by $\alpha_C^{c}=\sin^{-1}[k_C/k_A]$.
Beyond such critical angle of incidence, we should replace $\alpha_{_C}$ by $[-i\chi]$
in the wave function as well as in the probability current density factor. Hence it is modified as
\begin{eqnarray}
 A_{C}^{e}=1/\sqrt{(a_{C}^{e}-b_{C}^{e}c_{C}^{e})(e^{-\chi}+e^{-\chi^{*}})}\ .
\end{eqnarray}

Similarly, the critical angles corresponding to the other reflection channels can also be obtained following the same way.
The wave vectors at other reflection channels are $|k_{B}|=|k_{A}|=[\sqrt{(\eps+\mu_L)^2-\omega^2}+U]/(\hbar v_F)$,
$|k_{A'}|=|k_C|$ and $|k_{D(F)}|=[U\pm\sqrt{(\mu_L-\eps)^2-\omega^2}]/(\hbar v_F)$. The other coefficients are 
$b_{i}^{e(h)}=[(\hbar v_F|k_i|)^2+\omega^2-(U-\mu_N\mp\eps)^2]/(2\omega U)$,
$c_{i}^{e(h)}=[\omega-b_{i}^{e(h)}(U+\mu_L\pm\eps)]/(\hbar v_F|k_i|)$ and 
$a_{i}^{e(h)}=[\hbar v_F|k_i|b_{i}^{e(h)}+c_{i}^{e(h)}(U+\mu_L\pm\eps)]/\omega$.

Inside the superconducting region, the wave function will be similar to the case of NS junction~\cite{PhysRevB.93.195404}
except the appearance of four additional components. It is because of the fact that in NSN junction, Cooper pair inside 
the \suc can also be formed by pairing with an electron from the right normal region too. The wave function in the S-region can be written as
\begin{equation}
 \Psi_{S}=\sum_{\varrho=\pm,\mu=1}^2\mathcal{T}_{\mu}^{\varrho}e^{i\varrho k_{\mu}x}
 \left[\begin{array}[c]{c}1\\\mathcal{A}_{\mu}\\\mathcal{B}_{\mu}\\\mathcal{C}_{\mu}\\\mathcal{D}_{\mu}\\\mathcal{F}_{\mu}\\\mathcal{G}_{\mu}
 \\\mathcal{H}_{\mu}\end{array}\right]
 +\sum_{\varrho=\pm, \nu=3}^4\mathcal{T}_{\nu}^{\varrho}e^{-i\varrho k_{\nu}x}
 \left[\begin{array}[c]{c}1\\\mathcal{A}_{\nu}\\\mathcal{B}_{\nu}\\\mathcal{C}_{\nu}\\\mathcal{D}_{\nu}\\\mathcal{F}_{\nu}\\
 \mathcal{G}_{\nu}\\\mathcal{H}_{\nu}\end{array}\right]
\end{equation}
where $k_\mu=k_{0\mu}+i\kappa_\mu$ and $k_\nu=k_{o\nu}-i\kappa_\nu$. Here, $k_{\mu}$ and $k_{\nu}$ are evaluated using Eq.(\ref{band_S}). 
Also, $\mathcal{T}_{\mu}^{\varrho}$ and $\mathcal{T}_{\nu}^{\varrho}$ are the scattering amplitudes inside the superconducting region.
All the other coefficients are same as provided in Ref. [\onlinecite{PhysRevB.93.195404}] like
\begin{equation}
 \mathcal{A}_{\mu(\nu)}=\frac{(\varLambda_3\varLambda_8)^2-\mathcal{P}_{\mu(\nu)}\mathcal{N}_{\mu(\nu)}
 -\varLambda_{2l}\varLambda_{10}\varLambda_{4\mu(\nu)}\varLambda_{11}}
 {\varLambda_{2\mu(\nu)}(\mathcal{P}_{\mu(\nu)} \varLambda_{11}+\mathcal{N}_{\mu(\nu)}\varLambda_{10})}.
\end{equation}
\begin{equation}
 \mathcal{B}_{\mu(\nu)}=\frac{\mathcal{N}_{\mu(\nu)}+\varLambda_{2\mu(\nu)}\varLambda_{11}A_{\mu(\nu)}}{\varLambda_3\varLambda_8},
\end{equation}
\begin{equation}
 \mathcal{C}_{\mu(\nu)}=\frac{\mathcal{A}_{\mu(\nu)}\mathcal{N}_{\mu(\nu)}+\varLambda_{4\mu(\nu)}\varLambda_{11}}{\varLambda_3\varLambda_8},
\end{equation}
\begin{eqnarray}
 \mathcal{H}_{\mu(\nu)}=-\varLambda_3+\varLambda_5\mathcal{B}_{\mu(\nu)}+\varLambda_{2\mu(\nu)}\mathcal{C}_{\mu(\nu)}\ , \\ 
 \mathcal{G}_{\mu(\nu)}=\varLambda_{3}\mathcal{A}_{\mu(\nu)}-\varLambda_{4\mu(\nu)}\mathcal{B}_{\mu(\nu)}-\varLambda_{5}\mathcal{C}_{\mu(\nu)}\ , \\
 \mathcal{D}_{\mu(\nu)}=\varLambda_{4\mu(\nu)}+\varLambda_{1}\mathcal{A}_{\mu(\nu)}+\varLambda_{3}\mathcal{C}_{\mu(\nu)}\ , \\
 \mathcal{F}_{\mu(\nu)}=-\varLambda_{1}-\varLambda_{2\mu(\nu)}\mathcal{A}_{\mu(\nu)}-\varLambda_{3}\mathcal{B}_{\mu(\nu)}\ .
\end{eqnarray}
with~~~
$\varLambda_{3}=\omega/\Delta_{S}$,~~~$\varLambda_{6(7)}=[U\mp(\mu_S-\eps)]/\Delta_{S}$, \\
$\varLambda_{8(9)}=2\mu_S/\Delta_{S}$~~$\rm{and}$~~$\varLambda_{10(11)}=2(U\pm\mu_S)/\Delta_{S}$, \\
$\varLambda_{1(5)}=[U\mp(\mu_S+\eps)]/\Delta_{S}$,~~~$\varLambda_{2\mu}=\varLambda^{\ast}_{4\mu}=(k_y+ik_\mu)/\Delta_{S}$,\\
$\varLambda_{2\nu}=\varLambda^{\ast}_{4\nu}=(k_y-ik_\nu)/\Delta_{S}$.

and
\begin{eqnarray}
N_{\mu(\nu)}=\varLambda_1\varLambda_6+\varLambda_{2\mu(\nu)}\varLambda_{4\mu(\nu)}+\varLambda_3^2+1\ , \\
P_{\mu(\nu)}=\varLambda_5\varLambda_7+\varLambda_{2\mu(\nu)}\varLambda_{4\mu(\nu)}+\varLambda_3^2+1 \ .
\end{eqnarray}

The wave  function for the $p$ doped right normal region corresponding to the doping concentration $\mu_{_R}=-1.5\mu_{_L}$ is given by

\begin{widetext}
\begin{eqnarray}
 \Psi_{R}&=&\sum_{\bar{\iota}=G,H}A_{\bar{\iota}}^{e}t_{\bar{\iota}}^{e}e^{-ik_{\bar{\iota}}^{x}(x+L)}\left[\begin{array}[c]{c}
               1\\
               ia_{\bar{\iota}}^{e}e^{-i\alpha_{\bar{\iota}}}\\
               b_{\bar{\iota}}^{e}\\
               ic_{\bar{\iota}}^{e}e^{-i\alpha_{\bar{\iota}}}\\0\\0\\0\\0\\
              \end{array}\right]
              +
              \sum_{\bar{\eta}=Q,J}A_{\bar{\eta}}^{h}t_{\bar{\eta}}^{h}e^{ik_{\bar{\eta}}^{x}(x+L)}\left[\begin{array}[c]{c}
               0\\0\\0\\0\\
               1\\
               -ia_{\bar{\eta}}^{h}e^{i\alpha_{\bar{\eta}}}\\
               b_{\bar{\eta}}^{h}\\
               -ic_{\bar{\eta}}^{h}e^{i\alpha_{\bar{\eta}}}\\
              \end{array}\right].
\end{eqnarray}
\end{widetext}
Similarly, the wave function in the right normal region for the $\mu_{_R}=-\mu_{_L}$ case can be expressed as
\begin{widetext}
\begin{eqnarray}
 \Psi_{R}&=&A_R^{e}t_{R}^{e}e^{ik_{R}^{x}(x+L)}\left[\begin{array}[c]{c}
               1\\
               -ia_{R}^{e}e^{i\alpha_R}\\
               b_R^{e}\\
               -ic_R^{e}e^{i\alpha_R}\\0\\0\\0\\0\\
              \end{array}\right]
              +A_S^{e}t_S^{e}e^{-ik_{S}^{x}(x+L)}\left[\begin{array}[c]{c}
               1\\
               ia_{S}^{e}e^{-i\alpha_S}\\
               b_S^{e}\\
               ic_S^{e}e^{-i\alpha_S}\\0\\0\\0\\0\\
              \end{array}\right]+
              \sum_{\bar{\eta}=Q,J}A_{\bar{\eta}}^{h}t_{\bar{\eta}}^{h}e^{ik_{\bar{\eta}}^{x}(x+L)}\left[\begin{array}[c]{c}
               0\\0\\0\\0\\
               1\\
               -ia_{\bar{\eta}}^{h}e^{i\alpha_{\bar{\eta}}}\\
               b_{\bar{\eta}}^{h}\\
               -ic_{\bar{\eta}}^{h}e^{i\alpha_{\bar{\eta}}}\\
              \end{array}\right].
\end{eqnarray}
\end{widetext}
\section{Expression of shot noise cross-correlation}
\label{App2}
Following Eq.(\ref{sij}) we can express the shot noise cross-correlation in terms of NR, AR, CAR and CT amplitudes as follows,
\begin{eqnarray}
S_{ij,\rm{A}}^{ee}(\epsilon)&=&-\frac{2e^2}{h}\Big[\Big(\left(t^e_{\rm H}(\epsilon)+t^e_{\rm Z}(\epsilon)\right)\left(r^{e~*}_{\rm B}(\epsilon)+r^{e~*}_{\rm C}(\epsilon)\right)\Big.\Big.\nonumber\\
&&\Big.\Big.+(t^{e~*}_{\rm H}(\epsilon)+t^{e~*}_{\rm Z}(\epsilon))(r^e_{\rm B}(\epsilon)+r^e_{\rm C}(\epsilon))\Big)^2 \Big. \nonumber \\
&&+\Big.\Big(\left(t^h_{\rm J}(\epsilon)+t^h_{\rm Q}(\epsilon)\right)\left(r^{h~*}_{\rm F}(\epsilon)r^{h~*}_{\rm D}(\epsilon)\right)\Big. \nonumber \\
&&+\Big.(t^{h~*}_{\rm J}(\epsilon)+t^{h~*}_{\rm Q}(\epsilon))(r^h_{\rm F}(\epsilon)+r^h_{\rm D}(\epsilon))\Big)^2\Big]\ ,
\end{eqnarray}
and
\begin{eqnarray}
S_{ij,\rm{A}}^{eh}(\epsilon)&=&-\frac{2e^2}{h}\Big[\Big(\left(t^h_{\rm J}(\epsilon)+t^h_{\rm Q}(\epsilon)\right)\left(r^{e~*}_{\rm B}(\epsilon)+r^{e~*}_{\rm C}(\epsilon)\right)\Big.\Big.\nonumber\\
&&\Big.\Big.+(t^{e~*}_{\rm H}(\epsilon)+t^{e~*}_{\rm Z}(\epsilon))(r^h_{\rm F}(\epsilon)+r^h_{\rm D}(\epsilon))\Big) \Big. \nonumber \\
&&\Big.\Big(\left(t^{h~*}_{\rm J}(\epsilon)+t^{h~*}_{\rm Q}(\epsilon)\right)\left(r^e_{\rm B}(\epsilon)+r^e_{\rm C}(\epsilon)\right)\Big. \nonumber \\
&&+\Big.(t^e_{\rm H}(\epsilon)+t^e_{\rm Z}(\epsilon))(r^{h~*}_{\rm F}(\epsilon)+r^{h~*}_{\rm D}(\epsilon))\Big)\Big]\ .
\end{eqnarray}
These expressions are valid for the incoming channel $\rm A$. Similar expressions can be written for the other channel $\rm A^{\prime}$ corresponding 
to which the transmission and reflection channels will be modified accordingly. Therefore, the total shot noise reads,
\begin{eqnarray}
S_{ij}(\epsilon)&=&\sum_{\eta} \Big[S_{ij,\eta}^{ee}(\epsilon)+S_{ij,\eta}^{eh}(\epsilon) \Big]
\end{eqnarray}
where, $S_{ij,\eta}^{ee}$ and $S_{ij,\eta}^{eh}$ represent the cross-correlation corresponding to the phenomenon of CT and CAR respectively when the incoming electron
is incident from channel $\rm {\eta}$. As the particle-hole symmetry is preserved, we can write
\begin{eqnarray}
S_{ij,\eta}^{ee}(\epsilon)&=&S_{ij,\eta}^{hh}(\epsilon)\ , \nonumber \\
S_{ij\eta}^{eh}(\epsilon)&=&S_{ij,\eta}^{he}(\epsilon) \ .
\end{eqnarray}
Using the above relations, we numerically compute the shot noise cross-correlation which we have explained in the main text.

\end{appendix}
\vspace{0.5cm}
\acknowledgments{PD acknowledges Department of Science and Technology (DST), India for the financial support through SERB NPDF (File 
no. PDF/2016/001178). We acknowledge Arun M. Jayannavar for his support and encouragement.}
\bibliography{bibfile}{}

\end{document}